\documentclass[11pt]{article}

\usepackage{jheppub}
\usepackage{url,amsmath,amscd,amsfonts,graphicx,marginnote,amssymb,mathtools}
\usepackage{amsthm}
\usepackage[all]{xy}
\usepackage{xcolor}
\usepackage{braket}
\usepackage{fancyhdr}

\newtheoremstyle{stuff}{\topsep}{\topsep}%
     {\itshape}
     {}
     {\bfseries}
     {}
     {.5em}
     {\thmnote{#3}}

\newtheorem{thm}{\textbf{Theorem}}
\numberwithin{thm}{section}

\numberwithin{cor}{section}
\newtheorem{lem}[thm]{\textbf{Lemma}}
\newtheorem{prop}[thm]{\textbf{Proposition}}

\theoremstyle{definition}

\newtheorem{rem}{Remark}

\title{Supereigenvalue Models and Topological Recursion}

\author[a]{Vincent Bouchard}
\author[b]{and Kento Osuga}

\affiliation[a]{Department of Mathematical \& Statistical Sciences\\
University of Alberta\\
632 CAB, Edmonton, Alberta T6G 2G1\\
Canada}

\affiliation[b]{Department of Physics\\
University of Alberta\\
4-181 CCIS, Edmonton, Alberta, T6G 2E1\\
Canada}

\emailAdd{vincent.bouchard@ualberta.ca}
\emailAdd{osuga@ualberta.ca}

\abstract{
We show that the Eynard-Orantin topological recursion, in conjunction with simple auxiliary equations, can be used to calculate all correlation functions of supereigenvalue models.}

\begin{document}

\maketitle

\section{Introduction}

It is well known that Hermitian matrix models satisfy Virasoro constraints. 
As a consequence of Virasoro constraints, correlation functions for Hermitian matrix models obey a system of equations, known as loop equations. In turn, these loop equations can be solved recursively to calculate all correlation functions of Hermitian matrix models. In fact, this recursive system can be generalized beyond matrix models \cite{CEO, EO, EO2, AMM}. The resulting abstract recursive formalism has become known in the literature as the Eynard-Orantin topological recursion.\footnote{In this paper, we will refer to the abstract recursive system that solves the loop equations in matrix models as the Eynard-Orantin topological recursion, even though the method for solving loop equations recursively in the particular context of matrix models pre-dates the abstract formulation of Eynard and Orantin.}
 It can be used to calculate enumerative invariants in a variety of contexts, such as Gromov-Witten invariants and Hurwitz numbers (see for instance \cite{BKMP,BM,DBOSS,EMS,EO3,FLZ2,FLZ3,Ma} and references therein). 

An interesting question arises: can we generalize this triumvirate of Virasoro constraints, loop equations, and topological recursion to the supersymmetric realm? A natural starting point is supereigenvalue models, originally introduced in \cite{AlvarezGaume:1991jd} -- see also \cite{AP,AlvarezGaume:1992mm,Becker:1992rk,CHMS, CHMS2, CHJMS,Kroll,McArthur:1993hw,Plefka1,Plefka2,Plefka:1996tt,Itoyama}. Those were constructed precisely so that they satisfy super-Virasoro contraints. It was also shown that correlation functions in supereigenvalue models obey super-loop equations. Our objective is to complete the triad and recursively solve the super-loop equations.

Our main result is that in fact, there is no need to introduce a ``super-topological recursion''. Indeed, the Eynard-Orantin topological recursion, in conjunction with simple auxiliary equations, is sufficient to calculate all correlation functions of supereigenvalue models. This is perhaps unexpected. But as we will show, it follows because supereigenvalue models are highly constrained. The free energy of supereigenvalue models is at most quadratic in the Grassmann parameters, and in fact it is completely determined in terms of the free energy of Hermitian matrix models \cite{Becker:1992rk,McArthur:1993hw}. It is for this reason that the Eynard-Orantin topological recursion is sufficient to solve the super-loop equations. 

However, the structure of the auxiliary equations hints at a new interpretation of the recursive structure in terms of super-geometry. Indeed, the Eynard-Orantin topological recursion starts with the geometry of a spectral curve. For supereigenvalue models, the starting point is also a spectral curve; that of the corresponding Hermitian matrix models. However, it must be supplemented with a Grassmann-valued polynomial equation, which can be thought of as a super-partner of the spectral curve. Together they form a ``super spectral curve'' (see for instance \cite{CHMS,CHMS2,CHJMS}). However it is not clear how to reformulate the topological recursion and the auxiliary equations in terms of a single recursive structure living on the super spectral curve. We leave this for future work.

This paper is organized as follows. We first review Hermitian matrix models, Virasoro constraints, loop equations and the Eynard-Orantin topological recursion in section 2. This section is meant to be a (hopefully) pedagogical introduction to these topics, from the viewpoint of formal Hermitian matrix models. Then, we turn to the study of supereigenvalue models in section 3. We show that the Eynard-Orantin topological recursion, in conjunction with auxiliary equations, is sufficient to calculate all correlation functions of supereigenvalue models in section 4. We also consider the super-Gaussian model as a simple example of the recursive formalism. 
We end with a brief discussion in section 5.

\acknowledgments

We would like to thank Piotr Su\l kowski for interesting discussions, and Jeffrey Kroll for collaboration in the initial stages of this project. This work is partly supported by an NSERC Discovery Grant.


\section{Hermitian Matrix Models and Topological Recursion}\label{sec:Hermitian matrix models}
In this section we introduce formal Hermitian matrix models, and review the connections between Virasoro constraints, loop equations and topological recursion. A detailed discussion can be found in \cite{Eynard:2016yaa}.

\subsection{Formal Hermitian Matrix Models}
The question of convergence of matrix integrals is a rather complex one. However, for many applications of matrix models in physics and enumerative geometry, convergence is not really necessary. More precisely, in this context --- even though it is not always explicitly mentioned --- we are often interested in so-called \emph{formal matrix models}, rather than convergent matrix models. The difference between the two is well explained in \cite{Eynard:2016yaa}.

In this paper we focus on formal matrix models. An important consequence of formal matrix models is that the quantities of interest, such as the partition function, the free energy and correlation functions, all possess a well-defined $1/N$ expansion. Let us now define formal matrix models.

\subsubsection{Partition Function and Free Energy}
Let $H_N$ be the space of Hermitian $N\times N$ matrices and $M\in H_N$. 
The partition function of a formal Hermitian 1-matrix model is given by the formal series
\begin{equation}
Z(t,t_3,\cdots,t_d;T_2;N)=\prod_{k=3}^d\sum_{n_k\geq0}\int_{H_N}dM\frac{1}{n_k!}\left(\frac{N}{t}\frac{t_k}{k}\text{Tr}(M^k)\right)^{n_k}e^{-\frac{NT_2}{2t}\text{Tr}(M^2)},\label{formal Z}
\end{equation}
where\footnote{It will become clear later on why we introduce the parameter $T_2$ here.} $T_2\neq0$ and the measure $dM$ is the $U(N)$ invariant Lebesgue measure
on $H_N$
\begin{equation}
dM=\frac{1}{2^{N/2}(\pi t/NT_2)^{N^2/2}}\prod_{i=1}^NdM_{ii}\prod_{i<j}d\text{Re}M_{ij}~ d\text{Im}M_{ij}.\label{measure}
\end{equation}
The normalization of the measure \eqref{measure} is chosen such that the partition function is exactly one when all coupling constants $t_k=0$. The free energy is then defined as
\begin{equation}
F(t,t_3,\cdots,t_d;T_2;N)=\log Z(t,t_3,\cdots,t_d;T_2;N).\label{free energy}
\end{equation}

In contrast, the partition function of a convergent Hermitian 1-matrix model is defined with the order of summation and integral in \eqref{formal Z} switched:
\begin{eqnarray}
Z_{{\rm conv}}(t,t_3,\cdots,t_d;T_2;N)&=&\int_{H_N}dM\prod_{k=3}^d\sum_{n_k\geq0}\frac{1}{n_k!}\left(\frac{N}{t}\frac{t_k}{k}\text{Tr}(M^k)\right)^{n_k}e^{-\frac{NT_2}{2t}\text{Tr}(M^2)}\nonumber\\
&=&\int_{H_N}dMe^{-\frac{N}{t}{\rm Tr}V(M)},\label{conv Z}
\end{eqnarray}
where
\begin{equation}
V(M)=\frac{T_2}{2}M^2-\sum_{k=3}^d\frac{t_k}{k}M^k\label{potential}
\end{equation}
is called the potential.
Since summation and integral do not commute in general, formal matrix models \eqref{formal Z} are different from convergent matrix models \eqref{conv Z}.

Our interest in this paper is only in formal Hermitian 1-matrix models. For simplicity, however, we often omit the arguments $(t,t_k;T_2;N)$, and also denote a formal Hermitian matrix model by
\begin{equation}
Z\overset{\mathrm{formal}}{=}\int_{H_N}dMe^{-\frac{N}{t}{\rm Tr}V(M)},
\end{equation}
with the understanding that the summation should be taken outside of the integral.

Hermitian matrix models possess a $U(N)$ gauge symmetry, $M \to U^{\dagger}MU$, where $U$ is an $N\times N$ unitary matrix. If we fix the gauge freedom such that $M$ is diagonalized, the partition function is given up to normalization by 

\begin{equation}
Z \propto \int\prod_{i=1}^Nd\lambda_i\Delta(\lambda)^2e^{-\frac{N}{t}\sum_{i=1}^NV(\lambda_i)},\label{diagonalized Z}
\end{equation}
where $\Delta(\lambda)=\prod_{i<j}^N(\lambda_i-\lambda_j)$ is the Vandermonde determinant.

\subsubsection{Correlation Functions}
As usual we define the expectation value of a function $f$ by
\begin{equation}
\langle {\rm Tr}f(M)\rangle=\frac{1}{Z}\int_{H_N} dM {\rm Tr}( f(M))e^{-\frac{N}{t}{\rm Tr}V(M)},\label{def of expectation value}
\end{equation}
and we denote by $\langle \text{Tr} f(M) \rangle_{c}$ the corresponding connected expectation value.
We are interested in the expectation values:
\begin{equation}
T_{l_1\cdots l_n}(t,t_k;T_2;N)=\Bigl<{\rm Tr}(M^{l_1})\cdots{\rm Tr}(M^{l_n})\Bigr>_c.\label{expectation value}
\end{equation}
It turns out to be convenient to collect all $T_{l_1\cdots l_b}(t,t_k;T_2;N)$ for every nonnegative integers $l_1,\cdots,l_n$ in a single expression. We define the following generating functions, known as correlation functions:
\begin{eqnarray}
W_n(t, t_k; T_2;N; x_1,\cdots,x_n)&=&\sum_{l_1,\cdots,l_n\geq0}\frac{T_{l_1\cdots l_n}(t,t_k;T_2;N)}{x_1^{l_1+1}\cdots x_n^{l_n+1}},\label{W_n(J)}\nonumber\\
&=&\biggl<\prod_{j=1}^n{\rm Tr}\left(\frac{1}{x_j-M}\right)\biggr>_c.
\end{eqnarray}
The last equality is often used as a definition of the correlation functions; these should be understood as generating series in the variables $1/x_i$.

\subsubsection{$1/N$ Expansion}

For formal matrix models the free energy and correlation functions have a nice $1/N$ expansion. It follows from the definition of the partition function \eqref{formal Z} that the free energy \eqref{free energy} has an expansion
\begin{equation}
F(t,t_k;T_2;N)=\sum_{g\geq0}\left(\frac{N}{t}\right)^{2-2g}F_{g}(t,t_k;T_2),
\end{equation}
where the $F_{g}(t, t_k;T_2)$ do not depend on $N$. It can also be shown that the $F_{g}(t, t_k;T_2)$ are in fact power series in $t$ \cite{Eynard:2016yaa}.

A similar $1/N$ expansion also holds for correlation functions:
\begin{equation}
W_n(t, t_k; T_2;N; x_1,\cdots,x_n)=\sum_{g\geq0}\left(\frac{N}{t}\right)^{2-2g-n}W_{g,n}(t, t_k;T_2;x_1,\cdots,x_n),
\end{equation}
with the $W_{g,n}(t, t_k;T_2; x_1, \cdots, x_n)$ independent of $N$. Those are also power series in $t$. For simplicity of notation we will often drop the dependence on $t, t_k$ and $T_2$.

In fact, the expectation values $T_{l_1\cdots l_n}(t,t_k;T_2;N)$ can be interpreted in terms of ribbon graphs; we refer the reader to \cite{Eynard:2016yaa} for more details on this. It follows from the ribbon graph interpretation that they themselves have a $1/N$ expansion of the form:
\begin{equation}
T_{l_1\cdots l_n}(t,t_k;T_2;N)=\sum_{g\geq0}\left(\frac{N}{t}\right)^{2-2g-n}T^{(g)}_{l_1\cdots l_n}(t,t_k;T_2),
\end{equation}
thus we can write, order by order,
\begin{equation}
W_{g,n}(x_1,\cdots,x_n)=\sum_{l_1,\cdots,l_n\geq0}\frac{T^{(g)}_{l_1\cdots l_n}(t,t_k;T_2)}{x_1^{l_1+1}\cdots x_n^{l_n+1}}\label{W_{g,n}}.
\end{equation}
The $T^{(g)}_{l_1\cdots l_n}(t,t_k;T_2)$ are power series in $t$. Furthermore, it follows from the ribbon graph interpretation that if we collect the terms in the summation over $l_1, \cdots, l_n$ by powers of $t$, for each power of $t$ only a finite number of terms are non-zero. In other words, order by order in $t$, $W_{g,n}(x_1,\cdots,x_n)$ is polynomial in the variables $1/x_i$, $i=1,\cdots,n$ \cite{Eynard:2016yaa}.

\subsection{Virasoro Constraints and Loop Equations}
A fundamental result in the theory of formal matrix models is that Hermitian matrix models satisfy the so-called \emph{Virasoro constraints}. This implies a set of relations between correlation functions known as \emph{loop equations}.  Let us now review these properties of matrix models.

\subsubsection{Virasoro Constraints}\label{sec:Virasoro constraint}

In order to study the Virasoro constraints it is more convenient to extend the potential \eqref{potential} from a polynomial to a power series
\begin{equation}
V=\frac{T_2}{2} x^2 + \sum_{k\geq0}g_kx^k.\label{potential as a series}
\end{equation}
We will use this generalized potential to derive the Virasoro constraints and loop equations, but in the end we will set
\begin{equation}
g_k=-\frac{t_k}{k}\;\;(3\leq k\leq d),\;\;\;\;g_0=g_1=g_2 = g_k=0\;\;(k>d),\label{final interest}
\end{equation}
to recover a polynomial potential as in \eqref{potential}.

With this generalized potential we can obtain the correlation functions $W_{g,n}(x_1, \cdots, x_n)$ from the free energy by acting with the so-called \emph{loop insertion operator}, which is defined by
\begin{equation}
\frac{\partial}{\partial V(x)}=-\sum_{k\geq0}\frac{1}{x^{k+1}}\frac{\partial}{\partial g_k}.\label{loop insertion operator}
\end{equation}
Then $W_n(x_1,\cdots,x_n)$ and $W_{g,n}(x_1, \cdots, x_n)$ are obtained by:
\begin{eqnarray}
\label{loop insertion}
W_n(x_1, \ldots, x_n)&=&\left(\frac{N}{t}\right)^{-n}\frac{\partial}{\partial V(x_n)} \cdots \frac{\partial}{\partial V(x_1)} F,\label{generating function}\\
W_{g,n}(x_1, \ldots, x_n)&=&\frac{\partial}{\partial V(x_n)} \cdots \frac{\partial}{\partial V(x_1)} F_{g}.
\label{loop insertion genus}
\end{eqnarray}

We now define a sequence of operators $\{ L_n \}$, for $n \geq -1$:
\begin{equation}
L_n=T_2\frac{\partial}{\partial g_{n+2}} +  \sum_{k\geq0}kg_k\frac{\partial}{\partial g_{k+n}}+\frac{t^2}{N^2}\sum_{j=0}^n\frac{\partial}{\partial g_j}\frac{\partial}{\partial g_{n-j}}.\label{Virasoro generators}
\end{equation} 
Note that the third term is defined to be zero if $n=-1$. 
One can show that these operators are generators for the Virasoro subalgebra:
\begin{equation}
[L_m, L_n] = (m-n) L_{m+n}.\label{Witt}
\end{equation}

A fundamental result is that the formal Hermitian matrix model partition function with a generalized potential \eqref{potential as a series} is annihilated by the Virasoro operators. This is called the \emph{Virasoro constraints}: 
\begin{equation}
L_n Z=0, \qquad n \geq -1.\label{Virasoro constraint}
\end{equation}

\begin{rem}\label{uniqueness as solutions of the Virasoro constraint}
We could equivalently \emph{define} 1-cut Hermitian matrix models as solutions of the Virasoro constraints \eqref{Virasoro constraint}, as well as the additional constraint that
\begin{equation}
\frac{\partial Z}{\partial T_2}=\frac{1}{2}\frac{\partial Z}{\partial g_2}.
\end{equation}
It can be shown that these two constraints are sufficient to uniquely determine $Z$, which is none other than \eqref{formal Z}. See \cite{Alexandrov:2003pj,Alexandrov:2004ud,Alexandrov:2004ed} for more detail. We set $T_2=1$ for simplicity for the remainder of this section. 
\end{rem}

\subsubsection{Loop Equations}

From the Virasoro constraints satisfied by Hermitian matrix models one can derive a set of relations between correlation functions known as \emph{loop equations}. 

We start with the formal series in $1/x$:
\begin{eqnarray}
\label{loop equation 1}
0&=&\frac{1}{Z}\sum_{n\geq0}\frac{1}{x^{n+1}}L_{n-1}Z\nonumber\\
&=&\frac{1}{Z}\sum_{n\geq0}\frac{1}{x^{n+1}}\left(\frac{\partial}{\partial g_{n+1}}+\sum_{k\geq0}kg_k\frac{\partial}{\partial g_{k+n-1}}+\frac{t^2}{N^2}\sum_{j=0}^{n-1}\frac{\partial}{\partial g_j}\frac{\partial}{\partial g_{n-j-1}}\right)Z,
\end{eqnarray}
where the first equality holds due to the Virasoro constraints. We can rewrite \eqref{loop equation 1} using the fact that correlation functions can be obtained by acting on the free energy with the loop insertion operator \eqref{loop insertion}. After some manipulations we obtain the loop equation 
\begin{equation}
\label{eq:loop1}
-\frac{N}{t}V'(x)W_1(x)+P_1(x)+\Bigl(W_1(x)\Bigr)^2+W_2(x,x)=0,
\end{equation}
with
\begin{equation}
P_1(x)=- \frac{\partial}{\partial g_0} F -\sum_{m\geq0}x^m\sum_{k\geq0} (m+k+2)g_{m+k+2}\frac{\partial}{\partial g_k}F,
\end{equation}
where $V'(x)$ denotes the derivative  of the potential with respect to $x$. See Appendix~\ref{sec:Loop Equation for Hermitian Matrix Models} for the derivation.

Further, by acting an arbitrary number of times with the loop insertion operator on the loop equation, one obtains the general loop equation:
\begin{eqnarray}
\frac{N}{t}V'(x)W_{n+1}(x,J)&=&\sum_{I \subseteq J}W_{|I|+1}(x,I)W_{n-|I|+1}(x,J\backslash I)+W_{n+2}(x,x,J) \nonumber \\ &&+\sum_{i=1}^{n}\frac{\partial}{\partial x_i}\frac{W_{n}(x,J\backslash x_i)-W_n(J)}{x-x_i}+P_{n+1}(x,J),\label{general loop equation}
\end{eqnarray}
where we introduced the notation $J=(x_1,...,x_n)$, and $P_{n+1}(x;J)$ is defined by
\begin{equation}
P_{n+1}(x,J)=-\frac{t^n}{N^n}\left(\prod_{j=1}^{n}\frac{\partial}{\partial V(x_j)}\frac{\partial}{\partial g_0}+ \sum_{m\geq0}x^m\sum_{k\geq0} (m+k+2)g_{m+k+2}\prod_{j=1}^{n}\frac{\partial}{\partial V(x_j)}\frac{\partial}{\partial g_k} \right)F\label{def of P(x,J)}.
\end{equation}

If we insert the $1/N$ expansion in \eqref{general loop equation}, the coefficient of $(N/t)^{2-2g-n}$ gives the expansion of the loop equation:
\begin{eqnarray}
V'(x)W_{g,n+1}(x,J)&=&\sum_{I \subseteq J}\sum_{h=0}^gW_{h,|I|+1}(x,I)W_{g-h,n-|I|+1}(x,J\backslash I)+W_{g-1,n+2}(x,x,J) \nonumber \\
&&+\sum_{i=1}^{|J|}\frac{\partial}{\partial x_i}\frac{W_{g,n}(x,J\backslash x_i)-W_{g,n}(J)}{x-x_i}+P_{g,n+1}(x,J),\label{genus-expanded loop equation}
\end{eqnarray}
where $P_{g,n+1}(x,J)$ is defined by
\begin{equation}
P_{g,n+1}(x,J)=-\left(\prod_{j=1}^{n}\frac{\partial}{\partial V(x_j)}\frac{\partial}{\partial g_0}+ \sum_{m\geq0}x^m\sum_{k\geq0} (m+k+2)g_{m+k+2}\prod_{j=1}^{n}\frac{\partial}{\partial V(x_j)}\frac{\partial}{\partial g_k} \right)F_{g}\label{def of P_g(x,J)}.
\end{equation}

If we now set the potential to be a polynomial of degree $d$, that is, the coupling constants are chosen as in \eqref{final interest}, then the $P_{g,n+1}(x,J)$ defined in \eqref{def of P_g(x,J)} become polynomials in $x$ --- note that they are not necessarily polynomials with respect to $x_1, \cdots, x_n$ however. More precisely, $P_{0,1}(x)$ has degree $d-2$, while all other $P_{g,n+1}(x,J)$ are polynomials in $x$ of degree $d-3$. This is because we can rewrite
\begin{equation}
Z = e^{-\frac{N^2 g_0}{t}} \tilde Z, 
\end{equation}
where $\tilde Z$ does not depend on $g_0$, therefore
\begin{equation}
\frac{\partial F}{\partial g_0} = - \left(\frac{N}{t}\right)^2 t.
\label{eq:Fg0}
\end{equation}
Thus $\frac{\partial F_{0}}{\partial g_0} = -t$ while 
\begin{equation}
 \frac{\partial F_{g}}{\partial g_0} = 0 \qquad \text{for $g \geq 1$}.
\end{equation}
It then follows that the highest degree term $x^{d-2}$ in $P_{g,n+1}(x,J)$ is only non-vanishing for $(g,n) = (0,1)$.

\subsection{Topological Recursion}\label{sec:TR}
The loop equations \eqref{genus-expanded loop equation} provide a set of relations between correlation functions. However, each relation depends on a polynomial $P_{g,n+1}(x,J)$, see \eqref{def of P_g(x,J)}, which cannot \emph{a priori} be calculated from the matrix model. Fortunately, there is a method for recursively solving the loop equations for the correlation functions $W_{g,n}(x_1, \ldots, x_n)$, without first knowing the polynomials $P_{g,n+1}(x,J)$.

This recursive method can in fact be generalized beyond matrix models to the broader setup of algebraic geometry \cite{CEO, EO, EO2, AMM}. The resulting abstract recursive formalism has become known in the literature as the \emph{Eynard-Orantin topological recursion}. Henceforth we will refer to the recursive method for solving loop equations by this name as well. 

For completeness, let us review in this section the recursive method for solving loop equations in the context of formal Hermitian 1-matrix models.

\subsubsection{Planar Limit and Spectral Curve}
The loop equation \eqref{genus-expanded loop equation} for $g=0,n=1$ is
\begin{equation}
V'(x)W_{0,1}(x)=\bigl(W_{0,1}(x)\bigr)^2+P_{0,1}(x),
\end{equation}
or equivalently
\begin{equation}
\left(W_{0,1}(x)-\frac{1}{2}V'(x)\right)^2=\frac{1}{4}V'(x)^2-P_{0,1}(x).\label{W_{0,1} equation}
\end{equation}
Let us define
\begin{equation}\label{hyper}
y(x) = W_{0,1}(x) - \frac{1}{2}{V'(x)},
\end{equation}
so that \eqref{W_{0,1} equation} can be rewritten as
\begin{equation}\label{eq:hyper curve}
y(x)^2 = \frac{1}{4}V'(x)^2-P_{0,1}(x)
\end{equation}
The loop equation \eqref{genus-expanded loop equation} can also be rewritten in terms of $y(x)$. We obtain:
\begin{eqnarray}
- 2 y(x) W_{g,n+1}(x,J)&=&\sum_{I \subseteq J}^* \sum_{h=0}^gW_{h,|I|+1}(x,I)W_{g-h,n-|I|+1}(x,J\backslash I)+W_{g-1,n+2}(x,x,J) \nonumber \\
&&+\sum_{i=1}^{|J|}\frac{\partial}{\partial x_i}\frac{W_{g,n}(x,J\backslash x_i)-W_{g,n}(J)}{x-x_i}+P_{g,n+1}(x,J),\label{genus-expanded loop equation with y}
\end{eqnarray}
where $\displaystyle \sum_{I \subseteq J}^* \sum_{h=0}^g$ means that we are excluding the cases $(h,I)=(0, \emptyset)$ and $(h,I) = (g, J)$.
 
The 1-cut Brown's lemma \cite{Brown,Eynard:2006bs,DiFrancesco:1993cyw}, which applies to formal Hermitian 1-matrix models, implies that \eqref{eq:hyper curve} defines a (potentially singular) genus zero hyperelliptic curve of degree $2d-2$:

\begin{lem}\label{Brown's lemma}{\bf (1-cut Brown's Lemma)}
There exists a polynomial $M(x)$ of $x$ of degree $d-2$ whose roots $\alpha_i$ are power series of $t$, and a pair $a,b$ of power series of $\sqrt{t}$ where $a+b$ and $ab$ are power series of $t$, such that
\begin{equation}\label{hyperelliptic}
y^2 = M(x)^2 (x-a)(x-b),
\end{equation}
where
\begin{equation}
a=2\sqrt{t}+\mathcal{O}(t),\;\;\;\;b=-2\sqrt{t}+\mathcal{O}(t),\;\;\;\;\alpha_i=\alpha_i^0+\mathcal{O}(t),
\end{equation}
with non-zero constants $\alpha_i^0$.
\end{lem}

A key point here is that while everything so far was defined as formal series in $t$, in \eqref{hyperelliptic} all the $t$-dependence is in $a$, $b$ and the $\alpha_i$. In fact, it follows from the 1-cut Brown's lemma that the coefficients of the degree $2d-2$ polynomial in $x$ on the right-hand-side of \eqref{hyperelliptic} have a well defined power series expansion in $t$. We can even go further, and ``re-sum'' the power series; that is, we think of the coefficients as Taylor expansions of actual functions of $t$. In other words, we think of \eqref{hyperelliptic} as defining a $t$-dependent family of (potentially singular) genus zero hyperelliptic curves of degree $2d-2$.

This hyperelliptic curve, which is called the \emph{spectral curve}, plays a fundamental role for the topological recursion. In fact, we will want to interpret the correlation functions $W_{g,n}(x_1, \cdots, x_n)$ as ``living" on the spectral curve. Let us be a little more precise.

Since \eqref{hyperelliptic} has genus zero, we can parameterize it with rational functions:
\begin{eqnarray}
x(z)&=&\frac{a+b}{2}+\frac{a-b}{4}\left(z+\frac{1}{z}\right)\label{x(z)},\\
y(z)&=&M(x(z))\,\frac{a-b}{4}\left(z-\frac{1}{z}\right)\label{y(z)},
\end{eqnarray}
where $z$ is a coordinate on the Riemann sphere.\footnote{We abuse notation slightly here and use $y(z)$ to define the meromorphic function on the Riemann sphere, while we previously used $y(x)$ to denote its formal $t$-expansion with polynomial coefficients in $1/x$.} We can think of $x: \mathbb{C}_\infty \to \mathbb{C}_\infty$ as a branched double covering. Its two simple ramification points are at $z = \pm 1$, which are the two simple zeros of the one-form
\begin{equation}
dx(z)=\frac{a-b}{4}\left(1-\frac{1}{z^2}\right)dz.\label{dx}
\end{equation}
The hyperelliptic involution that exchanges the two sheets of $x: \mathbb{C}_\infty \to \mathbb{C}_\infty$ is given by $z \mapsto \sigma(z) = 1/z$, with
\begin{equation}
x(\sigma(z) ) = x(z), \qquad y(\sigma(z)) = - y(z).
\end{equation}

\subsubsection{Multilinear Differentials and Pole Structure}\label{sec:Multilinear Differentials and Pole Structure}

We now want to understand the correlation functions as living on the spectral curve. More precisely, we \emph{define} new objects, $\omega_{g,n}(z_1, \ldots, z_n)$, which are multilinear differentials on the Riemann sphere, and functions of $t$. In other words, they are multilinear differentials on the spectral curve. For $g \geq 0$, $n \geq 1$, and $2g-2+n \geq 1$, we define them such that
\begin{equation}
\omega_{g,n}(z_1, \cdots, z_n)=W_{g,n}(x_1, \cdots, x_n )dx_1 \cdots dx_n,\label{converse2}
\end{equation}
where we defined $x_i := x(z_i)$. By this equality, we mean that the Taylor expansion near $t=0$ of the multilinear differential on the left-hand-side recovers the formal series of the correlation functions on the right-hand-side. For the two remaining cases, we define
\begin{equation}
\omega_{0,1}(z) =  y(z) dx(z) = \left( W_{0,1}(x(z))  - \frac{1}{2} V'(x(z)) \right) dx(z),
\end{equation}
and
\begin{equation}
\omega_{0,2}(z_1, z_2) = \left( W_{0,2}(x(z_1), x(z_2)) + \frac{1}{(x(z_1) - x(z_2) )^2} \right) dx(z_1) dx(z_2).
\end{equation}

The $\omega_{g,n}$ are now honest multilinear differentials on the spectral curve, so we can study their properties. The most important aspect for us will be their pole structure. Let us start with $\omega_{0,2}(z_1, z_2)$.

For $(g,n) = (0,2)$, after multiplying by $dx(z_1) dx(z_2)$ the loop equation \eqref{genus-expanded loop equation with y} reduces to
\begin{align}
\omega_{0,2}(z_1, z_2) =& \frac{dx(z_1) dx(z_2)}{2 y(z_1)}  \left( \frac{d}{d x(z_2)}\frac{2 y(z_2) + V'(x(z_1)) - V'(x(z_2))}{2 (x(z_1)-x(z_2) )}+P_{0,2}(x(z_1),x(z_2)) \right) \nonumber\\
& + \frac{dx(z_1) dx(z_2)}{2 (x(z_1) - x(z_2) )^2}
\label{equation for W_0(z,z_1)}
\end{align}
We notice that the first line is odd under the hyperelliptic involution $z_1 \mapsto \sigma(z_1)$, while the second line is even. Thus $\omega_{0,2}(z,z_1)$ satisfies
\begin{equation}
\omega_{0,2}(z_1,z_2)+\omega_{0,2}(\sigma(z_1),z_2)=\frac{dx(z_1)dx(z_2)}{(x(z_1)-x(z_2))^2}.\label{relation between two omega_{0,2}s}
\end{equation}

Let us now study the pole structure of $\omega_{0,2} (z_1, z_2)$ in $z_1$; since it is symmetric under $z_1 \leftrightarrow z_2$, the same result will be true for $z_2$. From \eqref{equation for W_0(z,z_1)}, we see that $\omega_{0,2}(z_1,z_2)$ can have poles at the zeros of $y(z_1)$, at coinciding points $z_1 = z_2$ and $z_1 = \sigma(z_2)$, and at poles of $x(z_1)$.

First, we note that $\omega(z_1, z_2)$ cannot have poles at poles of $x(z_1)$, since $P_{0,2}(x(z_1); x(z_2))$ is a polynomial in $x(z_1)$ of degree at most $d-3$.

Let us now consider the zeros of $y(z_1)$ that are roots of $M(x(z_1))$. From \eqref{equation for W_0(z,z_1)} $\omega_{0,2}(z_1, z_2)$ can have at most poles of the form $1/(x(z_1) - \alpha_i)$ there. But by the 1-cut Brown's lemma, we know that $\alpha_i = \alpha_i^0 + \mathcal{O}(t)$ with a non-zero constant $\alpha_i^0$. Therefore, if we do a Taylor expansion near $t=0$, the constant term would have the form
\begin{equation}
\frac{1}{x_1 - \alpha_i^0} = \sum_{j \geq 0 } \frac{(\alpha_i^0)^j}{x_1^{j+1}},
\end{equation}
where we used $x_1 = x(z_1)$ for clarity. It would then contribute an infinite series in $1/x_1$ for a fixed power of $t$, which contradicts the statement that $\omega_{0,2}(z_1, z_2)$ should recover a formal expansion in $t$ with coefficients that are polynomials in $1/x_1$. Therefore $\omega_{0,2}(z_1,z_2)$ cannot have poles at the roots of $M(x(z_1))$.

This argument does not work however for the ramification points, which are simple zeros of $y(z_1)$. However, $dx(z_1)$ also has a simple zero there, hence $\omega(z_1, z_2)$ does not have poles at the ramification points.

All that remains are the coinciding points $z_1 = z_2$ and $z_1 = \sigma(z_2)$. As $z_1 \to \sigma(z_2)$, $y(z_1) \to y(\sigma(z_2)) = - y(z_2)$, and the double pole of the first line in \eqref{equation for W_0(z,z_1)} cancels out with the double pole of the second line. It thus follows that the only pole of $\omega_{0,2}(z_1,z_2)$ is a double pole at $z_1 = z_2$.

In fact, there is a unique bilinear differential on the spectral curve with a double pole at $z_1 =z_2$, no other pole, and that satisfies \eqref{relation between two omega_{0,2}s}:
\begin{equation}
\omega_{0,2}(z_1, z_2) = \frac{dz_1 dz_2}{(z_1 - z_2)^2}.
\end{equation}
This is the normalized bilinear differential of the second kind, which can be uniquely defined for Riemann surfaces of arbitrary genus \cite{Fay}. The normalization is of course trivial here since the Riemann surface has genus zero.

Let us now study the multilinear differentials $\omega_{g,n+1}$ for $2g-2+n \geq 0$. We first show that
\begin{equation}
\omega_{g,n+1}(z,J)+\omega_{g,n+1}(\sigma(z),J)=0,\label{anti-symmetric involution}
\end{equation}
where $J = \{z_1, \cdots, z_n \}$. We will prove this by induction on $2g-2+n$. The base cases are $\omega_{0,3}$ and $\omega_{1,1}$ with $2g-2+n = 0$. 

For $\omega_{1,1}$, the loop equation \eqref{genus-expanded loop equation with y} can be rewritten in terms of differentials as
\begin{equation}
-2 y(z_0) dx(z_0) \omega_{1,1}(z_0) = - \omega_{0,2}(\sigma(z_0), z_0) + P_{1,1}(x(z_0)) dx(z_0)^2.
\end{equation}
The two terms on the right-hand-side are clearly invariant under $z_0 \mapsto \sigma(z_0)$, hence
\begin{equation}
\omega_{1,1}(z_0) + \omega_{1,1}(\sigma(z_0) ) = 0.
\end{equation}

As for $\omega_{0,3}$, \eqref{genus-expanded loop equation with y} can be rewritten as
\begin{multline}
- 2 y(z_0) dx(z_0) \omega_{0,3}(z_0, z_1, z_2) = - \omega_{0,2}(z_0, z_1) \omega_{0,2}(\sigma(z_0), z_2) - \omega_{0,2}(\sigma(z_0), z_1)\omega_{0,2}(z_0, z_2) \\
 + dx(z_0)^2 \Bigg( dx(z_1) \frac{d}{dx(z_1)} \frac{\omega_{0,2}(\sigma(z_1), z_2)}{x(z_0) - x(z_1)}  + dx(z_2) \frac{d}{dx(z_2)} \frac{\omega_{0,2}(\sigma(z_1), z_2)}{x(z_0) - x(z_2)} \\
  + P_{0,3}(x(z_0),x(z_1),x(z_2)) dx(z_1) dx(z_2) \Bigg).
\end{multline}
The first two terms on the right-hand-side are exchanged under $z_0 \mapsto \sigma(z_0)$, while the remaining terms on the right-hand-side are invariant. Therefore
\begin{equation}
\omega_{0,3}(z_0, z_1, z_2) + \omega_{0,3}(\sigma(z_0), z_1, z_2) = 0.
\end{equation}
We now prove \eqref{anti-symmetric involution} by induction. Assume that it is true for all $(g,n)$ such that $0 \leq 2g-2+n < k$. We show that it implies that it must be true for $2g-2+n = k$. Assuming the induction hypothesis, for $2g-2+n \geq 1$ we can rewrite \eqref{genus-expanded loop equation with y} in terms of differentials as
\begin{multline}
2 y(z_0) dx(z_0) \omega_{g,n+1}(z_0, J) = \sum_{I \subseteq J}^* \sum_{h=0}^g \omega_{h,|I|+1}(z_0, I) \omega_{g-h,n-|I|+1}(\sigma(z_0),J\backslash I) \\+ \omega_{g-1,n+2}(z_0,\sigma(z_0),J) 
+ dx(z_0)^2 \Bigg(\sum_{i=1}^{|J|} dx(z_i) \frac{d}{dx(z_i)}\frac{\omega_{g,n}(J)}{x(z_0)-x(z_i)}\\ -  P_{g,n+1}(x(z_0),\cdots, x(z_n) )dx(z_1) \cdots dx(z_n) \Bigg),
\label{loop}
\end{multline}
with $J = \{ z_1, \cdots, z_n \}$. The first summation is invariant under $z_0 \mapsto \sigma(z_0)$, and all other terms on the right-hand-side are also invariant. Therefore
\begin{equation}
\omega_{g,n+1}(z_0, J) + \omega_{g,n+1}(\sigma(z_0), J) = 0,\label{hyperelliptic involution}
\end{equation}
and, by induction, this must hold for all $(g,n)$ such that $2g-2+n \geq 0$.

Let us now study the pole structure for $\omega_{g,n+1}(z_0, J)$ in terms of $z_0$; since the correlation functions are symmetric the result will hold for all other $z_i$, $i=1,\cdots,n$ as well. The only possible poles are at zeros of $y(z_0)$, coinciding points $z_0 = z_i$ and $z_0 = \sigma(z_i)$, $i=1,\cdots,n$, and at poles of $x(z_0)$. First, there is no pole at poles of $x(z_0)$ since $P_{g,n+1}(x(z_0), x(z_1), \cdots, x(z_n))$ has degree $d-3$. Second, there is no pole at coinciding points $z_0\rightarrow z_i$ by the loop equation \eqref{genus-expanded loop equation with y} and no pole either at $z_0\rightarrow \sigma(z_i)$ by the anti-symmetric involution relation \eqref{anti-symmetric involution}. All that remains are zeros of $y(z_0)$. By the same argument as for $\omega_{0,2}(z_1,z_2)$, there cannot be poles at zeros of $M(x(z_0))$, otherwise the $\omega_{g,n+1}$ would have expansions in $t$ with coefficients that are not polynomials in $1/x(z_0)$. The only remaining possible poles are at the ramification points of $x: \mathbb{C}_\infty \to \mathbb{C}_\infty$, that is, $z_0 = \pm 1$. In contrast to $\omega_{0,2}(z_1,z_2)$, these poles can be of higher order, and $dx(z_0)$ is not sufficient to get rid of them.

\subsubsection{Topological Recursion}
 
We are now ready to solve the loop equations recursively to determine all $\omega_{g,n+1}$ from $\omega_{0,1}$ and $\omega_{0,2}$. Let us start with the loop equation \eqref{loop}, rewritten as
\begin{multline}
\omega_{g,n+1}(z_0, J ) =\frac{1}{2 \omega_{0,1}(z_0)} \Bigg( \sum_{I \subseteq J}^* \sum_{h=0}^g \omega_{h,|I|+1}(z_0, I) \omega_{g-h,n-|I|+1}(\sigma(z_0),J\backslash I) \\+ \omega_{g-1,n+2}(z_0,\sigma(z_0),J) \Bigg)
\\ + \frac{dx(z_0)}{2 y(z_0)}  \Bigg(\sum_{i=1}^{|J|} dx(z_i) \frac{d}{dx(z_i)}\frac{\omega_{g,n}(J)}{x(z_0)-x(z_i)} -  P_{g,n+1}(x(z_0),\cdots, x(z_n) )dx(z_1) \cdots dx(z_n) \Bigg) ,
\label{loop2}
\end{multline}
with $J = \{z_1, \ldots, z_n\}$.
It is clear that the third line of the expression has no pole at the ramification points in $z_0$. Thus, if we evaluate the residue of the expression on the right-hand-side at the ramification points, the third line does not contribute. We now take advantage of this fact to construct the so-called topological recursion.

Let us introduce the normalized differential of the third kind $\omega^{a-b}(z)$, which has simple poles at $z=a$ and $z=b$ with residues $+1$ and $-1$ respectively. It is given by
\begin{equation}
\omega^{a-b}(z) = \int^{a}_{z'=b} \omega_{0,2}(z', z) =  \frac{dz}{z-a} - \frac{dz}{z-b}.
\end{equation}
This object can in fact be defined for Riemann surfaces of arbitrary genus as the integral (in the fundamental domain) of the normalized bilinear differential of the second kind.

Let $\alpha$ be a generic base point on the Riemann sphere, and consider $\omega^{z-\alpha}(z')$. While it is a one-form in $z'$, we can also think of it as a function in $z$. (Note however that this is only true on the Riemann sphere, on higher genus Riemann surfaces as a function of $z$ it is only defined in the fundamental domain.) It then follows that
\begin{equation}
\sum_{a \in \text{all poles}} \underset{w = a}{\text{Res }} \omega^{w-\alpha}(z_0) \omega_{g,n+1}(w, J) = 0.
\end{equation}
For $2g-2+n \geq 0$, the only poles of the integrand are at $w=z_0$ and at the ramification points $w = \pm 1$. The residue at $w=z_0$ gives
\begin{equation}
 \underset{w = z_0}{\text{Res }} \omega^{w-\alpha}(z_0) \omega_{g,n+1}(z_0, J)  = - \omega_{g,n+1}(z_0,J).
\end{equation}
It then follows that
\begin{equation}
\omega_{g,n+1}(z_0,J) = \sum_{a = \pm 1}  \underset{w = a}{\text{Res }}\omega^{w-\alpha}(z_0) \omega_{g,n+1}(w, J),
\end{equation}
and, substituting \eqref{loop2} in the right-hand-side, we obtain the topological recursion:
\begin{multline}
\omega_{g,n+1}(z_0,J) = \sum_{a \in \{-1, 1\}}  \underset{w = a}{\text{Res }} \frac{\omega^{w-\alpha}(z_0) }{2 \omega_{0,1}(w)} \Bigg( \sum_{I \subseteq J}^* \sum_{h=0}^g \omega_{h,|I|+1}(w, I) \omega_{g-h,n-|I|+1}(\sigma(w),J\backslash I) \\+ \omega_{g-1,n+2}(w,\sigma(w),J) \Bigg).
\label{eq:TR}
\end{multline}

This is a recursive formula which calculates all $\omega_{g,n+1}(z_0, J)$, $2g-2+n \geq 0$, from the initial data of a genus zero spectral curve 
\begin{equation}
y^2 = M(x) (x-a) (x-b),
\end{equation} 
a one-form
\begin{equation}
\omega_{0,1}(z) = y(z) dx(z) = \left( W_{0,1}(x(z)) - \frac{1}{2} V'(x(z)) \right) dx(z),
\end{equation}
and a bilinear differential
\begin{equation}
\omega_{0,2}(z_1, z_2) = \left( W_{0,2}(x(z_1), x(z_2)) + \frac{1}{(x(z_1) - x(z_2) )^2} \right) dx(z_1) dx(z_2).
\end{equation}

We note that the Eynard-Orantin topological recursion is of course much more general than this. It can be defined for (almost) arbitrary spectral curves, not just (singular) hyperelliptic genus zero curves \cite{EO,EO2,BE,BHLMR,AMM}; in fact, it was also recently reformulated in terms of quantization of Airy structures in \cite{KS,ABCD, Borot}. However, in the context of formal Hermitian 1-matrix models the formulation given here is sufficient to calculate all correlation functions.

\section{Supereigenvalue Models}

In this section we study supereigenvalue models. Those were introduced in \cite{AlvarezGaume:1991jd} and studied further in, for instance, \cite{AP,AlvarezGaume:1992mm,Becker:1992rk,CHMS, CHMS2, CHJMS,Kroll,McArthur:1993hw,Plefka1,Plefka2,Plefka:1996tt}.

The idea of supereigenvalue models is to construct a partition function that is annihilated by differential operators that are generators for a super-Virasoro subalgebra in the NS sector. The resulting partition function is not a matrix model, but it can be understood as a supersymmetric generalization of Hermitian matrix models in the eigenvalue formulation \eqref{diagonalized Z}, hence the name supereigenvalue models. 

From the super-Virasoro constraints one can also derive super-loop equations satisfied by correlation functions. The missing link then is whether there exists a recursive method for solving the super-loop equations. We show that, in fact, the standard Eynard-Orantin topological recursion, combined with simple auxiliary equations, is sufficient to calculate all correlation functions in supereigenvalue models.

\subsection{Supereigenvalue Models and Super-Virasoro Contraints}

Let us start by defining supereigenvalue models.

\subsubsection{Partition Function and Free Energy}
Let $V(x)$ be a power series potential \eqref{potential as a series}:
\begin{equation}
V(x) =\frac{T_2}{2}x^2 + \sum_{k \geq 0} g_k x^k,
\end{equation}
 and define a fermionic potential $\Psi(x)$ as
\begin{equation}
\Psi(x) = \sum_{k \geq 0} \xi_{k+\frac{1}{2}} x^k,
\end{equation}
where the $\xi_{k+\frac{1}{2}}$ are Grassmann coupling constants. 

We define the partition function of the \emph{formal supereigenvalue model} as
\begin{equation}
\mathcal{Z}(t_s,g_k,\xi_{k+\frac{1}{2}};T_2;2N)\overset{\mathrm{formal}}{=}\int d\lambda d\theta \Delta(\lambda, \theta)e^{-\frac{2N}{t_s}\sum_{i=1}^{2N}\bigl(V(\lambda_i)+\Psi(\lambda_i)\theta_i\bigr)},\label{Z_S}
\end{equation}
where the measure is
\begin{equation}
d\lambda=\prod_{i=1}^{2N}d\lambda_i,\;\;\;\;d\theta=\prod_{i=1}^{2N}d\theta_i,
\end{equation}
with the $\theta_i $ Grassmann variables.  $\Delta(\lambda,\theta)$ will be determined shortly. One should keep in mind here that this is a formal model, that is, the summation should be understood as being outside the integral. Similar to formal Hermitian matrix models, it can be shown that $\mathcal{Z}$ is given by a formal power series in $t_s$.

The free energy $\mathcal{F}$ for the supereigenvalue model is defined as usual by
\begin{equation}
\mathcal{F}(t_s,g_k,\xi_{k+\frac{1}{2}};T_2;2N)=\log \mathcal{Z}(t_s,g_k,\xi_{k+\frac{1}{2}};T_2;2N).
\end{equation}

\begin{rem}
We will denote objects in supereigenvalue models, such as partition function, free energy, and correlation functions, with curly letters $\mathcal{Z}$, $\mathcal{F}$ and $\mathcal{W}_n$ to differentiate them from their Hermitian counterparts.
\end{rem}

\subsubsection{Super-Virasoro Constraints}

We want the partition function above to be annihilated by a sequence of differential operators that are generators for a closed subalgebra of the $\mathcal{N}=1$ superconformal algebra in the Neveu-Schwarz (NS) sector. Let us first define such operators, and then show that we can uniquely fix $\Delta(\lambda,\theta)$ such that the partition function is annihilated by these operators.

We define the super-Virasoro operators $L_n$, $G_{n+\frac{1}{2}}$ for $n\geq-1$ as
\begin{eqnarray}
L_n&=&T_2\frac{\partial}{\partial g_{n+2}} + \sum_{k\geq0}kg_k\frac{\partial}{\partial g_{k+n}}+\frac{1}{2}\left(\frac{t_s}{2N}\right)^2\sum_{j=0}^n\frac{\partial}{\partial g_j}\frac{\partial}{\partial g_{n-j}}\nonumber\\
&&+\sum_{k\geq0}\left(k+\frac{n+1}{2}\right)\xi_{k+\frac{1}{2}}\frac{\partial}{\partial \xi_{n+k+\frac{1}{2}}}+\frac{1}{2}\left(\frac{t_s}{2N}\right)^2\sum_{j=0}^{n-1}\left(\frac{n-1}{2}-j\right)\frac{\partial}{\partial \xi_{j+\frac{1}{2}}}\frac{\partial}{\partial \xi_{n-j-\frac{1}{2}}}, \nonumber\\ \label{L in super Virasoro operators}\\ 
G_{n+\frac{1}{2}}&=&T_2\frac{\partial}{\partial \xi_{n + \frac{5}{2}}}+\sum_{k\geq0}\left(kg_k\frac{\partial}{\partial \xi_{n+k+\frac{1}{2}}}+\xi_{k+\frac{1}{2}}\frac{\partial}{\partial g_{k+n+1}}\right)+\left(\frac{t_s}{2N}\right)^2\sum_{j=0}^n\frac{\partial}{\partial \xi_{j+\frac{1}{2}}}\frac{\partial}{\partial g_{n-j}}\label{G in super Virasoro operators},
\end{eqnarray}
where $\Sigma^{-1}_{k=0},\Sigma^{-2}_{k=0}$ are defined to be zero. These operators are generators for the super-Virasoro subalgebra \cite{AlvarezGaume:1991jd} :
\begin{eqnarray}
&[L_m, L_n]&=(m-n)L_{m+n}, \nonumber \\
&\left[L_m, G_{n+\frac{1}{2}}\right]&= \left(\frac{m-1}{2}-n\right)G_{n+m+\frac{1}{2}} \label{super Virasoro subalgebra}, \\
&\Bigl\{G_{m+\frac{1}{2}}, G_{n+\frac{1}{2}}\Bigr\}&=2L_{n+m+1}. \nonumber
\end{eqnarray}

We now want to impose the super-Virasoro constraints, that is, we want the partition function $\mathcal{Z}$ to satisfy the requirement that
\begin{equation}
G_{n+\frac{1}{2}} \mathcal{Z}=0, \qquad L_n \mathcal{Z} = 0, \qquad n \geq -1.
\label{eq:sVc}
\end{equation}
First, we note that the condition $L_n \mathcal{Z} = 0$, $n \geq -1$, is automatically satisfied if $G_{n+\frac{1}{2} } \mathcal{Z} =0$, $n \geq -1$, by the super-Virasoro algebra \eqref{super Virasoro subalgebra}. So we only need to impose the fermionic condition.

It is straightforward to show that there is a unique choice of $\Delta(\lambda,\theta)$, up to overall rescaling, such that $G_{n+ \frac{1}{2}} \mathcal{Z} = 0$, $n \geq -1$:
\begin{equation}
\Delta(\lambda,\theta)=\prod_{i<j}^{2N}(\lambda_i-\lambda_j-\theta_i\theta_j).\label{measure of the supereigenvalue model}
\end{equation}
It is now clear why we defined the partition function \eqref{Z_S} with $2N$ eigenvalues; if the number of eigenvalues is odd, with the $\Delta(\lambda,\theta)$ above the eigenvalue integral becomes zero. Hence we must require an even number of eigenvalues. 

\subsubsection{Super-Virasoro Constraints and Free Energy}

The super-Virasoro constraints is the requirement that the partition function $\mathcal{Z}$ satisfies the equations
\begin{equation}
G_{n+\frac{1}{2}} \mathcal{Z}=0, \qquad L_n \mathcal{Z} = 0, \qquad n \geq -1.
\end{equation}
In this section we do formal manipulations of these equations to rewrite them in terms of the fermionic expansion of the free energy $\mathcal{F} = \log \mathcal{Z}$. This will be useful for us later on.

Let us do a power series expansion of $\mathcal{F}$ in the Grassmann coupling constants $\xi_{k+\frac{1}{2}}$. First, we know that only terms with an even number of Grassmann coupling constants will be non-vanishing in the expansion, since $\mathcal{F}$ is a bosonic quantity. We then introduce the notation
\begin{equation}
\mathcal{F} = \sum_{k\geq 0} \mathcal{F}^{(2 k)},
\end{equation}
where $\mathcal{F}^{(2 k)}$ denotes the term of order $2 k$ in the Grassmann coupling constants. For instance, $\mathcal{F}^{(2)}$ is quadratic in the $\xi_{k+\frac{1}{2}}$.

The condition $G_{n+\frac{1}{2}} \mathcal{Z}=0$, $n \geq -1$, rewritten in terms of the free energy $\mathcal{F}$, becomes
\begin{multline}
T_2\frac{\partial \mathcal{F}}{\partial \xi_{n + \frac{5}{2}}}+\sum_{k\geq0}\left(kg_k\frac{\partial \mathcal{F}}{\partial \xi_{n+k+\frac{1}{2}}}+\xi_{k+\frac{1}{2}}\frac{\partial \mathcal{F}}{\partial g_{k+n+1}}\right)\\
+\left(\frac{t_s}{2N}\right)^2\sum_{j=0}^n \left(\frac{\partial^2 \mathcal{F}}{\partial \xi_{j+\frac{1}{2}} \partial g_{n-j}}+ \frac{\partial \mathcal{F}}{\partial \xi_{j+\frac{1}{2}}}\frac{\partial \mathcal{F}}{\partial g_{n-j}}\right)= 0.
\end{multline}
Identifying terms by terms in the expansion in the Grassmann coupling constants, we get the system of equations
\begin{multline}
T_2\frac{\partial \mathcal{F}^{(2l)}}{\partial \xi_{n + \frac{5}{2}}}+\sum_{k\geq0}\left(kg_k\frac{\partial \mathcal{F}^{(2l)}}{\partial \xi_{n+k+\frac{1}{2}}}+\xi_{k+\frac{1}{2}}\frac{\partial \mathcal{F}^{(2l-2)}}{\partial g_{k+n+1}}\right)\\
+\left(\frac{t_s}{2N}\right)^2\sum_{j=0}^n  \left( \frac{\partial^2 \mathcal{F}^{(2l)}}{\partial \xi_{j+\frac{1}{2}} \partial g_{n-j}}+ \sum_{m=1}^l \frac{\partial \mathcal{F}^{(2m)}}{\partial \xi_{j+\frac{1}{2}}}\frac{\partial \mathcal{F}^{(2l-2m)}}{\partial g_{n-j}}\right)= 0,
\label{eq:FE1}
\end{multline}
for $l \geq 1$.

The other Virasoro constraints, $L_n \mathcal{Z} = 0$, $n \geq -1$, becomes, in terms of $\mathcal{F}$,
\begin{multline}
T_2\frac{\partial \mathcal{F}}{\partial g_{n+2}} + \sum_{k\geq0}kg_k\frac{\partial\mathcal{F}}{\partial g_{k+n}}+\frac{1}{2}\left(\frac{t_s}{2N}\right)^2\sum_{j=0}^n \left(\frac{\partial^2\mathcal{F}}{\partial g_j \partial g_{n-j}}+ \frac{\partial\mathcal{F}}{\partial g_j}\frac{\partial\mathcal{F}}{\partial g_{n-j}} \right)\\
+\sum_{k\geq0}\left(k+\frac{n+1}{2}\right)\xi_{k+\frac{1}{2}}\frac{\partial \mathcal{F}}{\partial \xi_{n+k+\frac{1}{2}}}\\
+\frac{1}{2}\left(\frac{t_s}{2N}\right)^2\sum_{j=0}^{n-1}\left(\frac{n-1}{2}-j\right)\left(\frac{\partial^2 \mathcal{F}}{\partial \xi_{j+\frac{1}{2}}\partial \xi_{n-j-\frac{1}{2}}}+   \frac{\partial \mathcal{F}}{\partial \xi_{j+\frac{1}{2}}}\frac{\partial \mathcal{F}}{\partial \xi_{n-j-\frac{1}{2}}} \right) = 0.
 \end{multline}
 Order by order in the Grassmann coupling constants, we get
 \begin{multline}
T_2\frac{\partial \mathcal{F}^{(2l)}}{\partial g_{n+2}} + \sum_{k\geq0}kg_k\frac{\partial\mathcal{F}^{(2l)}}{\partial g_{k+n}}+\frac{1}{2}\left(\frac{t_s}{2N}\right)^2\sum_{j=0}^n \left(\frac{\partial^2\mathcal{F}^{(2l)}}{\partial g_j \partial g_{n-j}}+ \sum_{m=0}^l \frac{\partial\mathcal{F}^{(2m)}}{\partial g_j}\frac{\partial\mathcal{F}^{(2l-2m)}}{\partial g_{n-j}} \right)\\
+\sum_{k\geq0}\left(k+\frac{n+1}{2}\right)\xi_{k+\frac{1}{2}}\frac{\partial \mathcal{F}^{(2l)}}{\partial \xi_{n+k+\frac{1}{2}}}\\
+\frac{1}{2}\left(\frac{t_s}{2N}\right)^2\sum_{j=0}^{n-1}\left(\frac{n-1}{2}-j\right)\left(\frac{\partial^2 \mathcal{F}^{(2l+2)}}{\partial \xi_{j+\frac{1}{2}}\partial \xi_{n-j-\frac{1}{2}}}+  \sum_{m=1}^{l} \frac{\partial \mathcal{F}^{(2m)}}{\partial \xi_{j+\frac{1}{2}}}\frac{\partial \mathcal{F}^{(2l+2-2m)}}{\partial \xi_{n-j-\frac{1}{2}}} \right) = 0,
\label{eq:FE2}
 \end{multline}
 for $l \geq 0$. 

\subsection{Quadratic Truncation and Relation to Hermitian Matrix Models}

Let us now come back to the formal supereigenvalue model. A remarkable fact, originally proven in \cite{McArthur:1993hw}, is that the free energy $\mathcal{F}$ of the formal supereigenvalue model contains the Grassman coupling constants $\xi_{k+\frac{1}{2}}$ only up to quadratic order. That is highly non-trivial. In the notation above, this means that
\begin{equation}
\mathcal{F} = \mathcal{F}^{(0)} + \mathcal{F}^{(2)}.\label{quadratic formula}
\end{equation}
For completeness, we provide a proof of this truncation for supereigenvalue models in Appendix~\ref{sec:Derivation of the Becker's Free Energy Formula}. 

\begin{rem}
While we have not investigated this closely yet, we do not expect the truncation of the free energy to quadratic order to hold in multi-cut supereigenvalue models; we expect non-vanishing higher order terms. This is because the proof is based on a careful permutation of indices of $\lambda_i,\theta_i$, which cannot be freely done in multi-cut models. In the formal language, this means that the truncation only holds for formal supereigenvalue models, in which case the spectral curve has genus zero. We do not expect it to hold for formal ``multi-supereigenvalue models'' for which the spectral curve would have higher genus.
\end{rem}

It turns out that this truncation of the fermionic expansion of $\mathcal{F}$ implies that $\mathcal{F}$ is closely related to the free energy of the formal Hermitian 1-matrix model $F$. More precisely, setting $t_s = 2 t$, we get the following relation, which was proven in  \cite{Becker:1992rk,McArthur:1993hw}:
\begin{prop}\label{p:becker}
\begin{equation}
\mathcal{F}(2t,g_k,\xi_{k+\frac{1}{2}};T_2;2N)=2\left(1-\sum_{k,l \geq 0}\xi_{k+\frac{1}{2}}\xi_{l+\frac{1}{2}}\frac{\partial^2}{\partial g_l\partial g_{k+1}}\right) F(t,g_k;T_2;N).\label{Beckers formula}
\end{equation}
Note that the free energy on the left-hand-side is for the formal supereigenvalue model, while the free energy on the right-hand-side is for the formal Hermitian model. In other words,
\begin{align}
\mathcal{F}^{(0)}(2t, g_k; T_2;2 N) =& 2 F(t, g_k;T_2; N)\\
\mathcal{F}^{(2)}(2t, g_k, \xi_{k + \frac{1}{2}}; T_2;2 N) =& -2 \sum_{k,l \geq 0}\xi_{k+\frac{1}{2}}\xi_{l+\frac{1}{2}}\frac{\partial^2}{\partial g_l\partial g_{k+1}}F(t,g_k;T_2;N).
\end{align}
\end{prop}

This relation is fundamental. What it says is that the free energy of the formal supereigenvalue model is completely determined in terms of the free energy of the formal Hermitian matrix model.

Let us now provide a proof of this formula from the super-Virasoro constraints. Our proof is different in flavour to the original one in \cite{Becker:1992rk}. It is purely algebraic; we show that the relation is a direct consequence of the super-Virasoro constraints, if the existence of a solution that is quadratic in the Grassman coupling constants is assumed.
\begin{proof}
Assume that
\begin{equation}
\mathcal{F} = \mathcal{F}^{(0)} + \mathcal{F}^{(2)},
\end{equation}
which is the case for the free energy of formal supereigenvalue models. \eqref{eq:FE1} for $l=2$ becomes
\begin{equation}
\sum_{k\geq0} \xi_{k+\frac{1}{2}}\frac{\partial \mathcal{F}^{(2)}}{\partial g_{k+n+1}}
+\left(\frac{t_s}{2N}\right)^2\sum_{j=0}^n  \frac{\partial \mathcal{F}^{(2)}}{\partial \xi_{j+\frac{1}{2}}}\frac{\partial \mathcal{F}^{(2)}}{\partial g_{n-j}}= 0.
\end{equation}
For $n=-1$, this is simply
\begin{equation}
\sum_{l \geq0} \xi_{l+\frac{1}{2}}\frac{\partial \mathcal{F}^{(2)}}{\partial g_{l}} = 0,
\label{eq:n-1}
\end{equation}
For $n=0$, we use the fact that $\mathcal{Z} = e^{- \frac{2N^2 g_0}{t}} \tilde{\mathcal{Z}}$, where $\tilde{\mathcal{Z}}$ does not depend on $g_0$, to see that $\mathcal{F}^{(2)}$ does not depend on $g_0$. Thus we get
\begin{equation}
\sum_{k\geq0} \xi_{k+\frac{1}{2}}\frac{\partial \mathcal{F}^{(2)}}{\partial g_{k+1}} = 0.
\label{eq:n0}
\end{equation}

On the one hand, \eqref{eq:n-1} means that
\begin{equation}
 \mathcal{F}^{(2)} = \sum_{l \geq0} \xi_{l+\frac{1}{2}}\frac{\partial \mathcal{A}^{(1)}}{\partial g_{l}}
\end{equation}
for some $ \mathcal{A}^{(1)}$ which is linear in the Grassmann parameters $\xi_{k + \frac{1}{2}}$. On the other hand, \eqref{eq:n0} says that
\begin{equation}
 \mathcal{F}^{(2)} = \sum_{k\geq0} \xi_{k+\frac{1}{2}}\frac{\partial \tilde{\mathcal{A}}^{(1)}}{\partial g_{k+1}}
\end{equation}
for some $\tilde{\mathcal{A}}^{(1)}$ that is also linear in the $\xi_{k+\frac{1}{2}}$. Therefore
\begin{equation}
\mathcal{F}^{(2)} = \sum_{k,l \geq 0}\xi_{k+\frac{1}{2}}\xi_{l+\frac{1}{2}}\frac{\partial^2 \hat F^{(0)}}{\partial g_l\partial g_{k+1}},
\label{eq:F2inF0}
\end{equation}
where $\hat F^{(0)} = \hat F^{(0)}(t,g_k;T_2;N)$ is some unknown function of $t$, $g_k$, $T_2$ and $N$, which is independent of the Grassmann parameters $\xi_{k+\frac{1}{2}}$. 

Let us now consider \eqref{eq:FE2} for $l=1$ and $n=0$. We get:
\begin{equation}
T_2\frac{\partial \mathcal{F}^{(2)}}{\partial g_{2}} + \sum_{k\geq0}kg_k\frac{\partial\mathcal{F}^{(2)}}{\partial g_{k}}
+\sum_{k\geq0}\left(k+\frac{1}{2}\right)\xi_{k+\frac{1}{2}}\frac{\partial \mathcal{F}^{(2)}}{\partial \xi_{k+\frac{1}{2}}}=0,
\end{equation}
where we used the fact that $\mathcal{F}^{(2)}$ is independent of $g_0$. Substituting \eqref{eq:F2inF0}, we get
\begin{align}
0 =& \sum_{k,l \geq 0}\xi_{k+\frac{1}{2}}\xi_{l+\frac{1}{2}} \left(T_2\frac{\partial^3 \hat F^{(0)}}{\partial g_l\partial g_{k+1}\partial g_2} + \sum_{m \geq 0} m g_m \frac{\partial^3 \hat F^{(0)}}{\partial g_l\partial g_{k+1}\partial g_{m}} + \left(k+l + 1\right) \frac{\partial^2 \hat F^{(0)}}{\partial g_l\partial g_{k+1}}  \right)\nonumber\\
=& \sum_{k,l \geq 0}\xi_{k+\frac{1}{2}}\xi_{l+\frac{1}{2}} \frac{\partial^2}{\partial g_l \partial g_{k+1}} \left(T_2 \frac{\partial \hat F^{(0)}}{\partial g_2} + \sum_{m \geq 0} m g_m \frac{\partial \hat F^{(0)}}{\partial g_{m}} \right).
\label{eq:g2} 
\end{align}
We will need this equation soon.

Let us now consider \eqref{eq:FE1} for $l=1$ and $n=-1$. We have
\begin{equation}
T_2\frac{\partial \mathcal{F}^{(2)}}{\partial \xi_{ \frac{3}{2}}}+\sum_{k\geq0}\left(\xi_{k+\frac{1}{2}}\frac{\partial \mathcal{F}^{(0)}}{\partial g_{k}}+kg_k\frac{\partial \mathcal{F}^{(2)}}{\partial \xi_{k-\frac{1}{2}}}\right)= 0.
\end{equation}
Substituting \eqref{eq:F2inF0}, we get
\begin{align}
0=&\sum_{l \geq 0}\xi_{l+\frac{1}{2}}\left(T_2 \frac{\partial^2 \hat F^{(0)}}{\partial g_l\partial g_{2}} - T_2\frac{\partial^2 \hat F^{(0)}}{\partial g_1\partial g_{l+1}}  +\frac{\partial \mathcal{F}^{(0)}}{\partial g_{l}} + \sum_{m \geq 0} m g_m \left(  \frac{\partial^2 \hat F^{(0)}}{\partial g_m \partial g_l} - \frac{\partial^2 \hat F^{(0)}}{\partial g_{m-1} \partial g_{l+1}} \right)  \right)\nonumber \\
=&\sum_{l \geq 0}\xi_{l+\frac{1}{2}} \left(\frac{\partial}{\partial g_l} \left(T_2 \frac{\partial \hat F^{(0)}}{\partial g_{2}}  + \sum_{m \geq 0} m g_m \frac{\partial \hat F^{(0)}}{\partial g_m} \right)  - \frac{\partial}{\partial g_{l+1}} \left(T_2\frac{\partial \hat F^{(0)}}{\partial g_1} + \sum_{m \geq 0} m g_m  \frac{\partial \hat F^{(0)}}{\partial g_{m-1}} \right) \right. \nonumber \\ 
&\quad  \left.   + \frac{\partial}{\partial g_{l}} (\mathcal{F}^{(0)} + \hat F^{(0)})  \right).
\end{align}
Let us now multiply by $\xi_{k+\frac{1}{2}}$ on the left, apply $\frac{\partial}{\partial g_{k+1}}$, and sum over $k$. We get:
\begin{equation}
 \sum_{k,l \geq 0} \xi_{k+\frac{1}{2}}\xi_{l+\frac{1}{2}}  \left(\frac{\partial^2}{\partial g_{k+1} \partial g_l} \left( T_2\frac{\partial \hat F^{(0)}}{\partial g_{2}}  + \sum_{m \geq 0} m g_m \frac{\partial \hat F^{(0)}}{\partial g_m} \right)   +  \frac{\partial^2}{\partial g_{k+1} \partial g_{l}} (\mathcal{F}^{(0)} + \hat F^{(0)}) \right)=0.
 \end{equation}
By \eqref{eq:g2}, the first term is zero. Therefore, we conclude that
\begin{equation}
\mathcal{F}^{(2)} = \sum_{k,l \geq 0} \xi_{k+\frac{1}{2}}\xi_{l+\frac{1}{2}} \frac{\partial^2}{\partial g_{k+1} \partial g_{l}}\hat F^{(0)} = - \sum_{k,l \geq 0} \xi_{k+\frac{1}{2}}\xi_{l+\frac{1}{2}} \frac{\partial^2}{\partial g_{k+1} \partial g_{l}}\mathcal{F}^{(0)}. 
\label{eq:F2F0}
\end{equation}

In other words, the free energy of the formal supereigenvalue model can be written as
\begin{equation}
\mathcal{F} = \left( 1 - \sum_{k,l \geq 0} \xi_{k+\frac{1}{2}}\xi_{l+\frac{1}{2}} \frac{\partial^2}{\partial g_{k+1} \partial g_{l}}\right) \mathcal{F}^{(0)}. \label{F^(2)}
\end{equation}
Note that so far we only used the super-Virasoro constraints for $n=-1$ and $n=0$.

What remains to be shown is that $\mathcal{F}^{(0)}(2 t, g_k; T_2;2N) = 2 F(t, g_k; T_2;N)$, where the right-hand-side is the free energy of the formal Hermitian matrix model. We go back to \eqref{eq:FE2} for $l=0$ and arbitrary $n$:
 \begin{multline}
T_2\frac{\partial \mathcal{F}^{(0)}}{\partial g_{n+2}} + \sum_{k\geq0}kg_k\frac{\partial\mathcal{F}^{(0)}}{\partial g_{k+n}}+\frac{1}{2}\left(\frac{t_s}{2N}\right)^2\sum_{j=0}^n \left(\frac{\partial^2\mathcal{F}^{(0)}}{\partial g_j \partial g_{n-j}}+ \frac{\partial\mathcal{F}^{(0)}}{\partial g_j}\frac{\partial\mathcal{F}^{(0)}}{\partial g_{n-j}} \right)\\
+\frac{1}{2}\left(\frac{t_s}{2N}\right)^2\sum_{j=0}^{n-1}\left(\frac{n-1}{2}-j\right)\left(\frac{\partial^2 \mathcal{F}^{(2)}}{\partial \xi_{j+\frac{1}{2}}\partial \xi_{n-j-\frac{1}{2}}}\right) = 0,
 \end{multline}
 We substitute \eqref{eq:F2F0}:
  \begin{multline}
T_2\frac{\partial \mathcal{F}^{(0)}}{\partial g_{n+2}} + \sum_{k\geq0}kg_k\frac{\partial\mathcal{F}^{(0)}}{\partial g_{k+n}}+\frac{1}{2}\left(\frac{t_s}{2N}\right)^2\sum_{j=0}^n \left(\frac{\partial^2\mathcal{F}^{(0)}}{\partial g_j \partial g_{n-j}}+ \frac{\partial\mathcal{F}^{(0)}}{\partial g_j}\frac{\partial\mathcal{F}^{(0)}}{\partial g_{n-j}} \right)\\
-\frac{1}{2}\left(\frac{t_s}{2N}\right)^2\sum_{j=0}^{n-1}\left(\frac{n-1}{2}-j\right)\left(\frac{\partial^2 \mathcal{F}^{(0)}}{\partial g_{n-j} \partial g_{j} }- \frac{\partial^2 \mathcal{F}^{(0)}}{\partial g_{j+1} \partial g_{n-j-1} } \right) = 0,
 \end{multline}
Using the fact that $\frac{\partial \mathcal{F}^{(0)}}{\partial g_0}$ is a constant, this simplifies to:
   \begin{equation}
T_2\frac{\partial \mathcal{F}^{(0)}}{\partial g_{n+2}} + \sum_{k\geq0}kg_k\frac{\partial\mathcal{F}^{(0)}}{\partial g_{k+n}}+\left(\frac{t_s}{2N}\right)^2\sum_{j=0}^n \left(\frac{\partial^2\mathcal{F}^{(0)}}{\partial g_j \partial g_{n-j}}+ \frac{1}{2}\frac{\partial\mathcal{F}^{(0)}}{\partial g_j}\frac{\partial\mathcal{F}^{(0)}}{\partial g_{n-j}} \right) = 0.
 \end{equation}
 Let us rewrite this equation in terms of $\tilde F(t,g_k;T_2;N) = \frac{1}{2} \mathcal{F}^{(0)} (2 t, g_k; T_2;2 N)$. We get
\begin{equation}
T_2\frac{\partial \tilde F}{\partial g_{n+2}} + \sum_{k\geq0}kg_k\frac{\partial \tilde F}{\partial g_{k+n}}+\left(\frac{t}{N}\right)^2\sum_{j=0}^n \left(\frac{\partial^2 \tilde F}{\partial g_j \partial g_{n-j}}+  \frac{\partial \tilde F}{\partial g_j}\frac{\partial \tilde F}{\partial g_{n-j}} \right) = 0,
\end{equation}
or equivalently in terms of $\tilde{Z}=e^{\hat{F}}$
\begin{equation}
T_2\frac{\partial \tilde Z}{\partial g_{n+2}} + \sum_{k\geq0}kg_k\frac{\partial \tilde Z}{\partial g_{k+n}}+\left(\frac{t}{N}\right)^2\sum_{j=0}^n\frac{\partial^2 \tilde Z}{\partial g_j \partial g_{n-j}} = 0.
\end{equation}
Furthermore, by the definition of supereigenvalue models \eqref{Z_S}, it is straightforward to obtain
\begin{equation}
\frac{\partial \tilde Z}{\partial T_2}=\frac{1}{2}\frac{\partial \tilde Z}{\partial g_2}.
\end{equation}
These two constraints are sufficient to determine that $\tilde F(t,g_k;T_2;N)$ is the free energy  of 1-cut formal Hermitian matrix models (see Remark \ref{uniqueness as solutions of the Virasoro constraint}). Thus, we conclude that $\tilde F(t,g_k;T_2;N) = F(t, g_k;T_2; N)$, that is, the free energy of the formal supereigenvalue model takes the form
\begin{equation}
\mathcal{F}(2t,g_k,\xi_{k+\frac{1}{2}};T_2;2N)=2\left(1-\sum_{k,l \geq 0}\xi_{k+\frac{1}{2}}\xi_{l+\frac{1}{2}}\frac{\partial^2}{\partial g_l\partial g_{k+1}}\right) F(t,g_k;T_2;N).
\end{equation}
\end{proof}

We now set $T_2=1$ for simplicity. With this under our belt, we can define the $1/N$ expansion of the free energy. Since $t_s=2t$, it is natural to define the $1/N$ expansion for $\mathcal{F}$ as
\begin{equation}
\mathcal{F}(2t,g_k,\xi_{k+\frac{1}{2}};2N)=\sum_{g\geq0}\left(\frac{N}{t}\right)^{2-2g}\mathcal{F}_{g}(2t,g_k,\xi_{k+\frac{1}{2}}).\label{1/N expansion of F_S}
\end{equation}
Then \eqref{Beckers formula} implies that
\begin{equation}
\mathcal{F}_{g}(2t,g_k,\xi_{k+\frac{1}{2}})=2\left(1-\sum_{k,l}\xi_{k+\frac{1}{2}}\xi_{l+\frac{1}{2}}\frac{\partial}{\partial g_{k+1}}\frac{\partial}{\partial g_l}\right)F_{g}(t,g_k).
\label{eq:beckN}
\end{equation}

\subsection{Correlation Functions}

Since the free energy of supereigenvalue models is completely determined in terms of the free energy of the Hermitian matrix model, we expect a similar statement to be true for correlation functions. 

The correlation functions of formal Hermitian 1-matrix models can be obtained by acting with the loop insertion operator \eqref{loop insertion operator} a number of times on the free energy, as shown in \eqref{generating function}. We can define correlation functions in supereigenvalue models in a similar way.

We define the following bosonic and fermionic loop insertion operators:
\begin{equation}
\frac{\partial}{\partial V(x)}=-\sum_{k\geq0} \frac{1}{x^{k+1}}\frac{\partial}{\partial g_k},\hspace{5mm} \frac{\partial}{\partial \Psi(X)}=-\sum_{k\geq0} \frac{1}{X^{k+1}}\frac{\partial}{\partial \xi_{k+\frac{1}{2}}}.
\end{equation}
Correlation functions are then obtained by:
\begin{eqnarray}
\mathcal{W}_{n|m}(J|K)&=&\left(\frac{N}{t}\right)^{-n-m}\prod_{j=1}^n\frac{\partial}{\partial V(x_j)}\prod_{i=1}^m\frac{\partial}{\partial \Psi(X_i)}\mathcal{F}(2t,g_k,\xi_{k+\frac{1}{2}};2N)\nonumber\\
&=&\sum_{k_1\cdots k_m\geq0}\sum_{l_1\cdots l_n\geq0}\sum_{a_1,\cdots a_m=1}^{2N}\sum_{b_1\cdots b_n=1}^{2N}\frac{\Bigl<\lambda_{a_1}^{k_1}\cdots\lambda_{a_n}^{k_n}\theta_{b_1}\lambda_{b_1}^{l_1}\cdots\theta_{b_m}\lambda_{b_m}^{l_m}\Bigr>_c}{x_1^{k_1+1}\cdots x_n^{k_n+1}X_1^{l_1+1}\cdots X_m^{l_m+1}},\label{def of (m,n)-point function by loop insertion operators}
\end{eqnarray}
where $J = \{x_1, \cdots, x_n\}$ and $K = \{ X_1, \cdots,  X_m \}$. We removed the dependence of the correlation functions on coupling constants for clarity.

As usual, the correlation functions inherit from \eqref{1/N expansion of F_S} a $1/N$ expansion:
\begin{equation}
\mathcal{W}_{n|m}(J|K) =\sum_{g \geq 0} \left( \frac{N}{t} \right)^{2-2g-m-n} \mathcal{W}_{g,n|m}(J|K).
\end{equation}
 We can further expand the correlation functions in terms of the fermionic coupling constants $\xi_{k+\frac{1}{2}}$. Since $\mathcal{F}$ is at most quadratic in the Grassmann parameters, \emph{i.e.} $\mathcal{F} = \mathcal{F}^{(0)} + \mathcal{F}^{(2)}$, we see that the only non-vanishing correlation functions have $0 \leq m \leq 2$. Further, we get
\begin{align}
\mathcal{W}_{g,n|0}(J|)=& \mathcal{W}^{(0)}_{g,n|0}(J|) + \mathcal{W}^{(2)}_{g,n|0}(J|),\\
\mathcal{W}_{g,n|1}(J|X_1) =& \mathcal{W}^{(1)}_{g,n|1}(J|X_1),\\
\mathcal{W}_{g,n|2}(J|X_1, X_2) =&\mathcal{W}^{(0)}_{g,n|2}(J|X_1, X_2),
\end{align}
where, as usual, the superscript denotes the terms of a given order in the Grassmann parameters.

Moreover, by \eqref{eq:beckN} we expect all these correlation functions to be somehow determined in terms of correlation functions of the Hermitian matrix model. For instance, it is clear that
\begin{equation}
\mathcal{W}^{(0)}_{g, n|0}( J|)= 2 W_{g,n}(J).
\end{equation}
Let us now study the other non-vanishing correlation functions.

\subsubsection{$\mathcal{W}^{(0)}_{g,n|2}(J|X_1, X_2)$}

We start with $\mathcal{W}^{(0)}_{g,n|2}(J|X_1, X_2)$. 
We have:
\begin{align}
\mathcal{W}^{(0)}_{g, n|2}(J|X_1, X_2 ) =&  \prod_{j=1}^n\frac{\partial}{\partial V(x_j)} \frac{\partial}{\partial \Psi(X_1)}\frac{\partial}{\partial \Psi(X_2)}\mathcal{F}^{(2)}_{g}(2t,g_k,\xi_{k+\frac{1}{2}}) \nonumber\\
=& - 2 \sum_{k, l \geq 0} \frac{1}{X_1^{k+1} X_2^{l+1}} \frac{\partial}{\partial \xi_{k+\frac{1}{2}}}  \frac{\partial}{\partial \xi_{l+\frac{1}{2}}}  \left( \sum_{i, j} \xi_{i+\frac{1}{2}} \xi_{j + \frac{1}{2}} \frac{\partial}{\partial g_{i+1}} \frac{\partial}{\partial g_{j}}\right) \prod_{j=1}^n\frac{\partial}{\partial V(x_j)} F_{g}(t, g_k)\nonumber\\
=& - 2 \sum_{k, l \geq 0} \frac{1}{X_1^{k+1} X_2^{l+1}} \left(\frac{\partial}{\partial g_{l+1}} \frac{\partial}{\partial g_k} - \frac{\partial}{\partial g_{k+1}} \frac{\partial}{\partial g_l} \right) W_{g,n}(J).
\label{eq:WS2}
\end{align}
We can simplify this further. Recall from \eqref{eq:Fg0} that
\begin{equation}
\frac{\partial F_0}{\partial g_0} = -  t, \qquad \frac{\partial F_g}{\partial g_0} = 0 \text{ for all $g \geq 1$.}
\end{equation}
Thus we can rewrite
\begin{align}
\sum_{k, l \geq 0}& \frac{1}{X_1^{k+1} X_2^{l+1}} \left(\frac{\partial}{\partial g_{l+1}} \frac{\partial}{\partial g_k} - \frac{\partial}{\partial g_{k+1}} \frac{\partial}{\partial g_l} \right) W_{g,n}(J)\nonumber\\
=& \sum_{k,l \geq 0} \left(\frac{1}{X_1^{k+1} X_2^{l}} - \frac{1}{X_1^{k} X_2^{l+1}} \right) \frac{\partial}{\partial g_{l}} \frac{\partial}{\partial g_{k}} W_{g,n}(J)\nonumber\\
=&(X_2 - X_1) \sum_{k,l \geq 0} \frac{1}{X_1^{k+1} X_2^{l+1}} \frac{\partial}{\partial g_{l}} \frac{\partial}{\partial g_{k}}W_{g,n}(J)\nonumber\\
=& (X_2 - X_1) \frac{\partial}{\partial V(X_1)}\frac{\partial}{\partial V(X_2)}W_{g,n}(J)\nonumber\\
=& (X_2 - X_1) W_{g,n+2}(X_1, X_2, J).
\end{align}
It thus follows that
\begin{equation}
\mathcal{W}^{(0)}_{g, n|2}(J|X_1, X_2 ) = 2 (X_1 - X_2) W_{g,n+2}(X_1, X_2, J).
\end{equation}

\subsubsection{$\mathcal{W}^{(1)}_{g,n|1}(J|X_1)$} 
Let us now turn to $\mathcal{W}^{(1)}_{g,n|1}(J|X_1)$. We do not need to do much work here. We note that
\begin{align}
\underset{X=\infty}{\text{Res }} \Psi(X) \mathcal{W}^{(0)}_{g,n|2}(J|X,X_1) d X =& \underset{X=\infty}{\text{Res }} \Psi(X) \frac{\partial}{\partial \Psi(X)} \mathcal{W}^{(1)}_{g,n|1}(J|X_1) d X \nonumber\\
=& - \underset{X=\infty}{\text{Res }}  \sum_{k \geq 0} \sum_{l \geq 0} \xi_{k + \frac{1}{2}} X^{k-l-1} \frac{\partial}{\partial \xi_{l + \frac{1}{2}}}  \mathcal{W}^{(1)}_{g,n|1}(J|X_1) d X\nonumber\\
=& \sum_{k \geq 0}  \xi_{k + \frac{1}{2}}\frac{\partial}{\partial \xi_{k + \frac{1}{2}}}  \mathcal{W}^{(1)}_{g,n|1}(J|X_1).
\end{align}
But since $ \mathcal{W}^{(1)}_{g,n|1}(J|X_1)$ depends linearly on the Grassmann coupling constants $\xi_{k + \frac{1}{2}}$, the operator $ \sum_{k \geq 0}  \xi_{k + \frac{1}{2}}\frac{\partial}{\partial \xi_{k + \frac{1}{2}}}$ is the identity operator. Hence, we get
\begin{equation}
\mathcal{W}^{(1)}_{g,n|1}(J|X_1) =  \underset{X=\infty}{\text{Res }} \Psi(X) \mathcal{W}^{(0)}_{g,n|2}(J|X,X_1) d X .\label{formula for W^1}
\end{equation}

\subsubsection{$\mathcal{W}^{(2)}_{g, n|0}(J|) $}
For $\mathcal{W}^{(2)}_{g, n|0}(J|) $, we get:
\begin{align}
\mathcal{W}^{(2)}_{g, n|0}(J|) =& \prod_{j=1}^n\frac{\partial}{\partial V(x_j)}\mathcal{F}^{(2)}_{g}(2t,g_k,\xi_{k+\frac{1}{2}}) \nonumber\\
=&  - 2 \sum_{k,l} \xi_{k+\frac{1}{2}}\xi_{l+\frac{1}{2}}\frac{\partial}{\partial g_{k+1}}\frac{\partial}{\partial g_l}  \prod_{j=1}^n\frac{\partial}{\partial V(x_j)} F_{g}(t, g_k)\nonumber\\
=& 2 \sum_{k,l} \xi_{k+\frac{1}{2}}\xi_{l+\frac{1}{2}}\frac{\partial}{\partial g_{k+1}}\frac{\partial}{\partial g_l}  W_{g,n}(J).
\label{eq:WS0}
\end{align}
We can use the same residue trick as for $\mathcal{W}^{(1)}_{g,n|1}(J|X_1)$. It then follows
\begin{equation}
\underset{X=\infty}{\text{Res }} \Psi(X) \mathcal{W}^{(1)}_{g,n|1}(J|X) d X =  \sum_{k \geq 0}  \xi_{k + \frac{1}{2}}\frac{\partial}{\partial \xi_{k + \frac{1}{2}}}  \mathcal{W}^{(2)}_{g,n|0}(J|).
\end{equation}
It is easy to see that the right-hand-side is $ 2\mathcal{W}^{(2)}_{g, n|0}(J |)$, and we obtain
\begin{eqnarray}
\mathcal{W}^{(2)}_{g, n|0}( J|) = \frac{1}{2}\underset{X=\infty}{\text{Res }} \Psi(X) \mathcal{W}^{(1)}_{g,n|1}(X|J) d X.
\end{eqnarray}

To summarize, we obtain the following relations between correlation functions for supereigenvalue models, which can be thought of as a consequence of Proposition \ref{p:becker} for correlation functions:
\begin{prop}\label{formulae for correlation functions}
\begin{align}
\mathcal{W}^{(0)}_{g, n|0}(J|) =& 2W_{g, n}(J) \nonumber\\
\mathcal{W}^{(0)}_{g, n|2}(J|X_1, X_2 ) =& 2 (X_1 - X_2) W_{g,n+2}(X_1, X_2, J),\nonumber\\
\mathcal{W}^{(1)}_{g,n|1}(J|X_1) =&  \underset{X=\infty}{{\rm Res }} \Psi(X) \mathcal{W}^{(0)}_{g,n|2}(J|X,X_1) d X,\nonumber\\
\mathcal{W}^{(2)}_{g, n|0}( J|) =& \frac{1}{2}\underset{X=\infty}{{\rm Res }} \Psi(X) \mathcal{W}^{(1)}_{g,n|1}(J|X) d X.
\label{eq:CRs}
\end{align}
\end{prop}

The important point here is that all correlation functions of formal supereigenvalue models are determined in terms of $W_{g,n}(J)$, the correlation functions of formal Hermitian matrix models.

\subsection{Super-Loop Equations}
Let us now turn to the study of super-loop equations. There are more than one type of loop equations in supereigenvalue models, depending on the order of the Grassmann coupling constants. We call loop equations with an even (resp. odd) dependence on the Grassmann parameters ``bosonic'' (resp. ``fermionic''). We simply give the equations here and leave their derivations to Appendix~\ref{sec:Fermionic Loop Equation for Supereigenvalue Models} and \ref{sec:Bosonic Loop Equation for Supereigenvalue Models}.

\subsubsection{Fermionic Loop Equation}
The derivation of the fermionic loop equation starts with the following formal series
\begin{equation}
\frac{1}{\mathcal{Z}}\sum_{n\geq0} \frac{1}{X^{n+1}}G_{n-\frac{1}{2}}\mathcal{Z}=0.\label{series for fermionic loop equation}
\end{equation}
After a few manipulations we obtain the fermionic loop equation:
\begin{equation}
-\frac{N}{t}V'(X)\mathcal{W}_{0|1}(|X)-\frac{N}{t}\Psi(X)\mathcal{W}_{1|0}(X|)+\mathcal{W}_{1|1}(X|X)+\mathcal{W}_{1|0}(X|)\mathcal{W}_{0|1}(|X)+\mathcal{P}_{0|1}(|X)=0,\label{fermionic loop equation}
\end{equation}
where
\begin{equation}
\mathcal{P}_{0|1}(|X)=\left(-\frac{\partial}{\partial \xi_{\frac{1}{2}}}-\sum_{k\geq0}X^k\left(\sum_{l\geq0}(k+l+2)g_{k+l+2}\frac{\partial}{\partial \xi_{l+\frac{1}{2}}}+\xi_{k+l+\frac{3}{2}}\frac{\partial}{\partial g_{l}}\right)\right)\mathcal{F}.
\end{equation}
Now we expand the fermionic loop equation \eqref{fermionic loop equation} in terms of $1/N$, and act an arbitrary number of times with the bosonic loop insertion operator on it. Collecting terms order by order in the Grassmann coupling constants, we get the following two fermionic loop equations:
\begin{multline}
V'(X)\mathcal{W}_{g,n|1}^{(1)}(J|X)+\Psi(X)\mathcal{W}_{g,n+1|0}^{(0)}(X,J|) -\mathcal{P}^{(1)}_{g,n|1}(J|X)\\
=\sum_{I\subseteq J}\sum_{h=0}^g \mathcal{W}_{h,m|1}^{(1)}(I|X)\mathcal{W}_{g-h,n-m+1|0}^{(0)}(X,J\backslash I|)+\mathcal{W}_{g-1,n+2}^{(1)}(X|X,J) \\
+\sum_{i=1}^{n}\frac{\partial}{\partial x_i}\frac{\mathcal{W}_{g,n-1|1}^{(1)}(J\backslash x_i|X)-\mathcal{W}_{g,n-1|1}^{(1)}(J\backslash x_i|x_i)}{X-x_i},\label{xi^1-equation w.r.t x}
\end{multline}
and
\begin{equation}
\Psi(X)\mathcal{W}_{g,n+1|0}^{(2)}(X,J|)-\mathcal{P}_{g,n|1}^{(3)}(J|X)=\sum_{I\subseteq J}\sum_{h=0}^g \mathcal{W}_{h,m+1|0}^{(2)}(X,I|)\mathcal{W}_{g-h,n-m|1}^{(1)}(J\backslash I|X),\label{xi^3-equation w.r.t x}
\end{equation}
where we defined
\begin{equation}
\mathcal{P}_{g,n|1}(J|X)=\left(-\frac{\partial}{\partial \xi_{\frac{1}{2}}}-\sum_{k\geq0}X^k\left(\sum_{l\geq0}(k+l+2)g_{k+l+2}\frac{\partial}{\partial \xi_{l+\frac{1}{2}}}+\xi_{k+l+\frac{3}{2}}\frac{\partial}{\partial g_{l}}\right)\right)\prod_{j=1}^n\frac{\partial}{\partial V(x_j)}\mathcal{F}_g\label{P_{g,n+1}^{(F,1)}(x|J)},
\end{equation}
which, by \eqref{eq:beckN}, has an expansion $\mathcal{P}_{g,n|1}(J|X)=\mathcal{P}^{(1)}_{g,n|1}(J|X)+\mathcal{P}^{(3)}_{g,n|1}(J|X)$. 
 
If we act with the fermionic loop insertion operator on \eqref{xi^1-equation w.r.t x}, we obtain the equation for $\mathcal{W}_{g,n|2}^{(0)}(J|X,X_1)$:
\begin{multline}
V'(X)\mathcal{W}_{g,n|2}^{(0)}(J|X,X_1)-\mathcal{P}^{(0)}_{g,n|2}(J|X,X_1) \\
 =\sum_{I\subseteq J}\sum_{h=0}^g \mathcal{W}_{h,m|2}^{(0)}(I|X,X_1)\mathcal{W}_{g-h,n-m|1}^{(0)}(J\backslash I|X)+\frac{\mathcal{W}_{g,n|1}^{(0)}(J|X)-\mathcal{W}_{g,n|1}^{(0)}(J|X_1)}{X-X_1}\\
 +\sum_{i=1}^{n}\frac{\partial}{\partial x_i}\frac{\mathcal{W}_{g,n-1|2}^{(0)}(J\backslash x_i|X,X_1)-\mathcal{W}_{g,n-1|2}^{(0)}(J\backslash x_i|x_i,X_1)}{X-x_i}+\mathcal{W}_{g-1,n+1|2}^{(0)}(X,J|X,X_1),\label{bar{xi}^0-equation w.r.t x}
\end{multline}
where 
\begin{equation}
\mathcal{P}^{(0)}_{g,n|2}(J|X,X_1)=\left(-\frac{\partial}{\partial \xi_{\frac{1}{2}}}-\sum_{k\geq0}X^k\sum_{l\geq0}(k+l+2)g_{k+l+2}\frac{\partial}{\partial \xi_{l+\frac{1}{2}}}\right)\prod_{j=1}^n\frac{\partial}{\partial V(x_j)}\frac{\partial}{\partial\Psi(X_1)}\mathcal{F}^{(2)}_g.
\end{equation}

\subsubsection{Bosonic Loop Equation}
To get the bosonic loop equation, we start with the formal series:
\begin{equation}
\frac{1}{\mathcal{Z}}\sum_{n\geq0} \frac{1}{x^{n+1}}L_{n-1}\mathcal{Z}=0.
\end{equation}
We manipulate the above equation to obtain the bosonic loop equation:
\begin{multline}
-\frac{N}{t}V'(x)\mathcal{W}_{1|0}(x|)+\frac{1}{2} \Bigl(\mathcal{W}_{1|0}(x|)\Bigr)^2+ \frac{1}{2} \mathcal{W}_{2|0}(x,x|)- \frac{N}{2 t} \Psi'(x) \mathcal{W}_{0|1}(|x) + \frac{N}{2 t} \Psi(x)\frac{\partial}{\partial x}  \mathcal{W}_{0|1}(|x)  \\
+\frac{1}{2} \mathcal{W}_{0|1}(|x)  \frac{\partial}{\partial x} \mathcal{W}_{0|1}(|x) +\frac{1}{2} \left. \frac{\partial}{\partial x} \left( \mathcal{W}_{0|2}(|x,x') \right) \right|_{x'=x} + \mathcal{P}_{1|0}(x|)= 0,\label{bosonic loop equation}
\end{multline}
where we defined
\begin{multline}
\mathcal{P}_{1|0}(x|) = -\frac{\partial  \mathcal{F}}{\partial g_0}-\sum_{n\geq0}x^n \left( \sum_{k\geq0} (n+k+2)g_{n+k+2}\frac{\partial \mathcal{F}}{\partial g_k} + \frac{1}{2} \sum_{k \geq 0}  \xi_{n+k+\frac{5}{2}}  \left( n+2 k+3\right) \frac{\partial \mathcal{F}}{\partial \xi_{k+\frac{1}{2}}} \right).
\end{multline}
Before we do a $1/N$ expansion, let us study the dependence on the Grassmann parameters. \eqref{Beckers formula} implies that the dependence of the bosonic loop equation is at most of order $4$. Since $\mathcal{P}_{1|0}(x|)$ depends on the Grassmann parameters at most quadratically, the order $4$ terms in the bosonic loop equation directly yield the condition:
\begin{equation}
\Bigl(\mathcal{W}^{(2)}_{1|0}(x|)\Bigr)^2=0.
\end{equation}
Let us now study the bosonic equation at order $2$ and $0$. We act an arbitrary number of times with the bosonic loop insertion operator on \eqref{bosonic loop equation}, and then do a $1/N$-expansion. We also collect terms according to their order in the Grassmann parameters. We obtain the two following equations:
\begin{multline}
V'(x)\mathcal{W}_{g,n+1|0}^{(0)}(x,J|)-\mathcal{P}^{(0)}_{g,n+1|0}(x,J|)\\
=\frac{1}{2}\sum_{I\subseteq J}\sum_{h=0}^g \mathcal{W}_{h,m+1|0}^{(0)}(x,I|)\mathcal{W}_{g-h,n-m+1|0}^{(0)}(x,J\backslash I|) +\frac{1}{2}\mathcal{W}_{g-1,n+2|0}^{(0)}(x,x,J|)\\
+\frac{1}{2}\frac{\partial}{\partial x}\mathcal{W}^{(0)}_{g-1,n|2}(J|x,x')\Bigr|_{x'=x}+\sum_{i=1}^{n}\frac{\partial}{\partial x_i}\frac{\mathcal{W}_{g,n|0}^{(0)}(x,J\backslash x_i|)-\mathcal{W}_{g,n|0}^{(0)}(J|)}{x-x_i},\label{xi^0-equation w.r.t x}
\end{multline}
and
\begin{multline}
V'(x)\mathcal{W}_{g,n+1|0}^{(2)}(x,J|)-\mathcal{P}^{(2)}_{g,n+1|0}(x,J|)\\
=\sum_{I\subseteq J}\sum_{h=0}^g\biggl(\mathcal{W}_{h,m+1|2}^{(2)}(x,I|)\mathcal{W}_{g-h,n-m+1|0}^{(0)}(x,J\backslash I|)-\frac{1}{2}\mathcal{W}_{h,m|1}^{(1)}(I|x)\frac{\partial}{\partial x}\mathcal{W}^{(1)}_{g-h,n-m|1}(J\backslash I|x)\biggr) \\
+\frac{1}{2}\mathcal{W}_{g-1,n+2|0}^{(2)}(x,x,J|)
+\frac{1}{2}\biggl(\Psi(x)\frac{\partial}{\partial x}\mathcal{W}^{(1)}_{g,n|1}(J|x)-\Psi'(x)\mathcal{W}_{g,n|1}^{(1)}(J|x)\biggr)\\
+\sum_{i=1}^{n}\frac{\partial}{\partial x_i}\frac{\mathcal{W}_{g,n|0}^{(2)}(x,J\backslash x_i|)-\mathcal{W}_{g,n|0}^{(2)}(J|)}{x-x_i},\label{xi^2-equation w.r.t x}
\end{multline}
where we defined
\begin{align}
\mathcal{P}^{(0)}_{g,n+1|0}(x,J|)=&\left(-\frac{\partial}{\partial g_0}-\sum_{l\geq0}x^l\sum_{k\geq0} (l+k+2)g_{l+k+2}\frac{\partial}{\partial g_k}\right)\prod_{j=1}^n\frac{\partial}{\partial V(x_j)}\mathcal{F}^{(0)}_g,\\
\mathcal{P}^{(2)}_{g,n+1|0}(x,J|)=&\Biggl(-\frac{\partial}{\partial g_0}-\sum_{l\geq0}x^l\Biggl(\sum_{k\geq0} (l+k+2)g_{l+k+2}\frac{\partial}{\partial g_k}\nonumber\\
&\quad +\frac{1}{2}\sum_{k\geq0}\xi_{l+k+\frac{5}{2}}(l+2k+3)\frac{\partial}{\partial \xi_{k+\frac{3}{2}}}\Biggr)\Biggr)\prod_{j=1}^n\frac{\partial}{\partial V(x_j)}\mathcal{F}^{(2)}_g.
\end{align}

To conclude this section, let us show that the bosonic loop equation that is independent of the Grassmann parameters, that is  \eqref{xi^0-equation w.r.t x}, is indeed equivalent to the loop equation for Hermitian matrix models \eqref{genus-expanded loop equation}. First of all, \eqref{eq:CRs} turns the third term on the right-hand-side in \eqref{xi^0-equation w.r.t x} into
\begin{equation}
\frac{1}{2}\frac{\partial}{\partial x}\mathcal{W}^{(0)}_{g-1,n|2}(J|x,x')\Bigr|_{x'=x}=\frac{\partial}{\partial x}\Bigl((x-x')W_{g-1,n+2}(x,x',J)\Bigr)\Bigr|_{x'=x}=W_{g-1,n+2}(x,x,J).
\end{equation}
Also, \eqref{eq:beckN} implies that $\mathcal{P}^{(0)}_{g,n+1|0}(x,J|)=2P_{g,n+1}(x,J)$. We further substitute $\mathcal{W}_{h,l|0}^{(0)}(L|)=2W_{h,l}(L)$ for all $h,l$ into \eqref{xi^0-equation w.r.t x} . We obtain
\begin{multline}
2V'(x)W_{g,n+1}(x,J)-2P_{g,n+1}(x,J)\\
=2\sum_{I\subseteq J}\sum_{h=0}^g W_{h,m+1}(x,I)W_{g-h,n-m+1}(x,J\backslash I) +W_{g-1,n+2}(x,x,J)\\
+W_{g-1,n+2}(x,x,J)+2\sum_{i=1}^{n}\frac{\partial}{\partial x_i}\frac{W_{g,n}(x,J\backslash x_i)-W_{g,n}(J)}{x-x_i},
\end{multline}
which is precisely twice the loop equation for Hermitian matrix models \eqref{genus-expanded loop equation}.

\section{Topological Recursion}

The correlation functions for supereigenvalue matrix models are fully determined by \eqref{eq:CRs}. What we do in this section is reformulate these equations in the geometric language of the Eynard-Orantin topological recursion.

To this end, we now restrict ourselves to polynomial potentials. That is, we choose the coupling constants $g_k$ such that $g_0=g_1=g_2 = 0$ and $g_j = 0$ for all $j > d$. For fermionic couplings, we set $\xi_{k + \frac{1}{2}} = 0$ for all $k > d'$. Note that $d$ and $d'$ are independent integers.

\subsection{Hermitian Mutilinear Differentials} 

Let us start by recalling what we did in Section~\ref{sec:TR} for Hermitian matrix models. We constructed a sequence of multilinear differentials $\omega_{g,n}(z_1, \ldots, z_n)$ on the Riemann sphere (and functions of $t$), such that, for $g \geq 0$, $n \geq 1$ and $2g-2+n \geq 1$:
\begin{equation}
\omega_{g,n}(z_1, \cdots, z_n)=W_{g,n}(x_1, \cdots, x_n )dx_1 \cdots dx_n,
\end{equation}
where $x_i := x(z_i)$.
By this equality, we meant that the Taylor expansion of the multilinear differential on the left-hand-side near $t=0$ recovers the formal series of the correlation functions on the right-hand-side.

We also defined the two ``unstable'' cases ($2g-2+n \leq 0$) as:
\begin{equation}
\omega_{0,1}(z) = \left( W_{0,1}(x(z)) - \frac{1}{2} V'(x(z))\right) dx(z)
\end{equation}
and
\begin{equation}
\omega_{0,2}(z_1,z_2) = \left( W_{0,2}(x(z_1),x(z_2)) + \frac{1}{(x(z_1)-x(z_2))^2} \right) dx(z_1) dx(z_2).
\end{equation}

Then we showed that:
\begin{enumerate}
\item There are two meromorphic functions $x(z)$ and $y(z)$ on the Riemann sphere such that $\omega_{0,1}(z) = y(z) dx(z)$ and
\begin{equation}
y^2 = M(x)^2 (x-a) (x-b),
\end{equation}
with $M(x)$ a polynomial of degree $d-2$. We call this hyperelliptic curve the \emph{spectral curve} of the matrix model. We generally choose the coordinate $z$ on the Riemann sphere as being given by the parameterization
\begin{eqnarray}
x(z)&=&\frac{a+b}{2}+\frac{a-b}{4}\left(z+\frac{1}{z}\right)\label{x(z)}, \nonumber\\
y(z)&=&M(x(z))\,\frac{a-b}{4}\left(z-\frac{1}{z}\right)\label{y(z)}
\label{eq:param}
\end{eqnarray}
of the hyperelliptic curve.

\item $\omega_{0,2}(z_1, z_2)$ takes a very simple form; it is the normalized bilinear differential of the second kind on the Riemann sphere, that is,
\begin{equation}
\omega_{0,2}(z_1, z_2) = \frac{d z_1 dz_2}{ (z_1-z_2)^2}.
\end{equation}

\item The multilinear differentials $\omega_{g,n}(z_1, \cdots, z_n)$, for $2g-2+n \geq 1$, satisfy the Eynard-Orantin topological recursion \eqref{eq:TR}. The initial conditions of the recursion are $\omega_{0,1}(z)$ and $\omega_{0,2}(z_1,z_2)$.
\end{enumerate}

\subsection{Supereigenvalue Multilinear Differentials}
We would like to extend these results to supereigenvalue correlation functions. More precisely, we would like to construct (Grassman-valued) multilinear differentials on the spectral curve in a similar way. For 
$2 g - 2 + m + n +p \geq 1$, we want to construct multilinear differentials:
\begin{equation}
\Omega^{(p)}_{g, n|m}(z_1, \cdots, z_n|w_1, \cdots, w_m) = \frac{1}{2} \mathcal{W}^{(p)}_{g, n|m} ( x_1, \cdots, x_n|X_1, \cdots, X_m) d x_1 \cdots dx _n d X_1 \cdots d X_m ,
\label{eq:MDwant}
\end{equation}
where $x_i := x(z_i)$ and $X_j := x(w_j)$. As usual, the equality here means that after Taylor-expanding the left-hand-side near $t=0$ we recover the formal series of the correlation functions on the right-hand-side. Note that the factor of $1/2$ is simply there for convenience.

For the ``unstable'' cases ($2g - 2 + m + n +p \leq 0$), we modify the definitions slightly as for Hermitian matrix models. We define:
\begin{align}
\Omega^{(0)}_{0, 1|0}(z|) =& \frac{1}{2} \left( \mathcal{W}^{(0)}_{0,1|0}(x(z)|) -  V'(x(z))\right) dx(z)\label{eq:O0010}\\
\Omega^{(0)}_{0, 2|0}(z_1, z_2|) =& \frac{1}{2} \left( \mathcal{W}^{(0)}_{0,2|0}(x(z_1),x(z_2)|) + \frac{1}{(x(z_1)-x(z_2))^2} \right) dx(z_1) dx(z_2)\\
\Omega^{(1)}_{0, 0|1}(|w) =& \frac{1}{2} \left( \mathcal{W}^{(1)}_{0,0|1}(|x(w)) -  \Psi(x(w))\right) dx(w)\label{eq:O1001}\\
\Omega^{(0)}_{0, 0|2}(|w_1, w_2) =&\frac{1}{2} \left( \mathcal{W}^{(0)}_{0,0|2}(|x(w_1),x(w_2)) + \frac{1}{x(w_1)-x(w_2)} \right) dx(w_1) dx(w_2).
\end{align}

Our goal is to show that we can construct such multilinear differentials.

\subsubsection{Spectral Curve}

As for Hermitian matrix models, we know that there exists two meromorphic functions $x(z)$ and $y(z)$ on the Riemann sphere such that $\Omega^{(0)}_{0,1|0}(z|) = y(z) dx(z)$ and 
\begin{eqnarray}
y^2 &=&\frac{1}{4}V'(x)^2-\frac{1}{2}\mathcal{P}_{0,1|0}^{(0)}(x|)\nonumber\\
&=&M(x)^2 (x-a)(x-b),\label{curve1 for SEM}
\end{eqnarray}
where $M(x)$ is a polynomial of degree $d-2$. We choose our coordinate $z$ on the Riemann sphere as in \eqref{eq:param}.

Next, we would like to define a Grassmann-valued meromorphic function $\gamma(z)$ on the Riemann sphere such that $\Omega^{(1)}_{0,0|1}(|w) = \gamma(w) dx(w)$. How is this function related to the spectral curve? Let us consider the fermionic loop equation \eqref{xi^1-equation w.r.t x} for $g=0,n=0$:
\begin{equation}
V'(x)\mathcal{W}_{0,0|1}^{(1)}(|x)-\mathcal{P}^{(1)}_{0,0|1}(|x)+\Psi(x)\mathcal{W}_{0,1|0}^{(0)}(x|) =\mathcal{W}_{0,0|1}^{(1)}(I|x)\mathcal{W}_{0,1|0}^{(0)}(x|).
\end{equation}
We rewrite it as
\begin{equation}
\Bigl(\mathcal{W}_{0,1|0}^{(0)}(x|)-V'(x)\Bigr)\Bigl(\mathcal{W}_{0,0|1}^{(1)}(|x)-\Psi(x)\Bigr)=V'(x)\Psi(x)-\mathcal{P}^{(1)}_{0,0|1}(|x).
\end{equation}
Using the definitions \eqref{eq:O0010} and \eqref{eq:O1001}, we get
\begin{equation}
y\gamma=P^{(1)}(x),\label{curve2 for SEM}
\end{equation}
where
\begin{equation}
P^{(1)}(x)=\frac{1}{4}\Bigl(V'(x)\Psi(x)-\mathcal{P}^{(1)}_{0,0|1}(|x)\Bigr)
\end{equation}
 is a Grassmann-valued polynomial of degree $d+d'-1$. In particular, note that \eqref{curve2 for SEM} implies that $\gamma(z)$ is odd under the hyperelliptic involution, that is, $\gamma(1/z) = - \gamma(z)$.
 
\eqref{curve2 for SEM} can be thought of as a superpartner to the spectral curve \eqref{curve1 for SEM}. Together they form a super spectral curve --- see \cite{CHMS, CHMS2, CHJMS} for more on this. Ultimately, it would be great to reformulate the recursive structure as living on this super spectral curve. We leave this for future work.

\subsubsection{Unstable Cases}
Let us consider the rest of the unstable multilinear differentials. As for Hermitian matrix models, we know from \eqref{eq:CRs} that $\Omega^{(0)}_{0,2|0}(z_1, z_2|)$ becomes the normalized bilinear differential of the second kind on the Riemann sphere:
\begin{equation}
\Omega^{(0)}_{0,2|0}(z_1, z_2|) = \omega_{0,2}(z_1, z_2)= \frac{d z_1 d z_2}{(z_1 - z_2)^2}.
\end{equation}
Accordingly, by \eqref{eq:CRs} we get
\begin{align}
\Omega^{(0)}_{0, 0|2}(|w_1, w_2) =& (x(w_1) - x(w_2)) \Omega^{(0)}_{0,2|0}(w_1, w_2|)\\
 =&   (x(w_1) - x(w_2)) \omega_{0,2}(w_1, w_2).
\end{align}

\subsubsection{Stable Cases}
\eqref{eq:CRs} of course plays a crucial role. Let $\{\omega_{g,n}(J)\}$ for $2g-2+n\geq1$ be the set of multilinear differentials for formal Hermitian matrix models obtained by the Eynard-Orantin topological recursion \eqref{eq:TR}. Then \eqref{eq:CRs}  implies that the stable multilinear differentials for $m=0,p=0$ and $m=2,p=0$ are determined by
\begin{eqnarray}
\Omega^{(0)}_{g,n|0}(J|)&=&\omega_{g,n}(J),\nonumber\\
\Omega^{(0)}_{g,n|2}(J|w_1,w_2)&=&(x(w_1)-x(w_2))\;\omega_{g,n+2}(w_1,w_2,J).\label{formula for Omega^0}
\end{eqnarray}
Note that as shown in \eqref{hyperelliptic involution} all stable $\omega_{g,n}(J)$ are odd under the hyperelliptic involution $z\mapsto\sigma(z)=1/z$. Thus, this must hold for stable $\Omega^{(0)}_{g,n|0}(J|)$ and $\Omega^{(0)}_{g,n|2}(J|w_1,w_2)$ as well. In particular,
\begin{equation}
\Omega^{(0)}_{g,n|2}(J|w_1,w_2)=-\Omega^{(0)}_{g,n|2}(J|\sigma(w_1),w_2)=-\Omega^{(0)}_{g,n|2}(J|w_1,\sigma(w_2)).
\end{equation}

We now turn to the case $m=1,p=1$. In order to obtain the differentials for this case, we would like to modify \eqref{eq:CRs} into a residue formula on the Riemann sphere. Notice that we can rewrite the third equation in \eqref{eq:CRs} as
\begin{equation}
\mathcal{W}^{(1)}_{g,n|1}(J|X_1) =  \underset{X=\infty}{\text{Res }} (\Psi(X)-\mathcal{W}^{(1)}_{0,0|1}(|X)) \mathcal{W}^{(0)}_{g,n|2}(J|X,X_1) d X,\label{formula for W^1 2}
\end{equation}
since $\mathcal{W}^{(1)}_{0,0|1}(|X) \mathcal{W}^{(0)}_{g,n|2}(J|X,X_1) d X$ is regular at $X \to \infty$.
Then, in terms of differentials on the Riemann sphere, and using the Grassmann-valued function $\gamma(z)$ introduced earlier, we obtain
\begin{equation}
\Omega^{(1)}_{g,n|1}(J|w_1) =- 2\underset{x(w)=\infty}{\text{Res }} \gamma(w) \Omega^{(0)}_{g,n|2}(J|w,w_1).
\end{equation}
Remark that the residue here still makes sense. This is because both $\gamma(w)$ and $\Omega^{(0)}_{g,n|2}(J|w,w_1)$ are odd under the hyperelliptic involution $w\rightarrow1/w$, hence the integrand itself is even, hence a well-defined differential form on the base $x(w)$.

We can rewrite this expression as a residue on the Riemann sphere itself:
\begin{equation}
2\underset{x(w)=\infty}{\text{Res}}\gamma(w)\Omega_{g,n|2}^{(0)}(J|w,w_1)=\underset{w = 0}{\text{Res}}\gamma(w)\Omega_{g,n|2}^{(0)}(J|w,w_1) + \underset{w = \infty}{\text{Res}}\gamma(w)\Omega_{g,n|2}^{(0)}(J|w,w_1).
\end{equation}
Finally, we notice that \eqref{formula for Omega^0} ensures that the integrand can have poles only at the ramification points of the $x$-covering (\emph{i.e.} at $w = \pm 1$) and at the poles of $x(w)$ (\emph{i.e.} $w = 0$ and $w = \infty$), since all stable $\omega_{g,n}(J)$ have poles only at the ramification points. Using the fact that the sum of all possible residues of a differential form on the Riemann sphere vanishes, we arrive at:
\begin{equation}
\Omega^{(1)}_{g,n|1}(J|w_1)= \sum_{a \in \{-1, 1\}} \underset{w=a}{\text{Res}}\ \gamma(w)\Omega_{g,n|2}^{(0)}(J|w,w_1),
\end{equation}
which holds for $2g+n\geq1$. In terms of the correlation functions of the Hermitian matrix model, we get
\begin{equation}
\Omega^{(1)}_{g,n|1}(J|w_1)= \sum_{a \in \{-1, 1\}} \underset{w=a}{\text{Res}}\ \gamma(w) (x(w)-x(w_1)) \omega_{g,n+2}(w, w_1, J).
\end{equation}

Following the same reasoning, we can turn the fourth equation in \eqref{eq:CRs} into a residue formula on the Riemann sphere. We obtain:
\begin{equation}
\Omega^{(2)}_{g,n|0}(J|)=\frac{1}{2}\sum_{a \in \{-1, 1\} }\underset{w=a}{\text{Res}}\ \gamma(w)\Omega_{g,n|1}^{(1)}(J|w),
\end{equation}
which is again valid for $2g+n \geq 1$. In terms of correlation functions of the Hermitian matrix model, we get
\begin{equation}
\Omega^{(2)}_{g,n|0}(J|)=\frac{1}{2}\sum_{a \in \{-1, 1\} } \sum_{b \in \{-1, 1\} } \underset{w=a}{\text{Res}}\  \underset{z=b}{\text{Res}}\left(\gamma(w) \gamma(z) (x(z)-x(w)) \omega_{g,n+2}(z,w,J) \right).
\end{equation}

Therefore, all correlation functions of supereigenvalue models can be determined using topological recursion on the spectral curve \eqref{curve1 for SEM}, in conjunction with auxiliary equations defined in terms of the Grassmann-valued polynomial equation \eqref{curve2 for SEM}. To summarize, we get:

\begin{thm}\label{main theorem}
Starting with the spectral curve \eqref{curve1 for SEM}, the Eynard-Orantin topological recursion constructs a sequence of multilinear differentials $\omega_{g,n}(J)$. Then the correlation functions of supereigenvalue models are encoded in the following Grassmann-valued multilinear differentials on the spectral curve \eqref{curve1 for SEM}. 

The unstable differentials are defined by
\begin{eqnarray}
\Omega^{(0)}_{0,1|0}(z_1|)&=&y(z_1)dx(z_1),\nonumber\\
\Omega^{(1)}_{0,1|0}(|z_1)&=&\gamma(z_1)dx(z_1),\nonumber\\
\Omega^{(0)}_{0,2|0}(z_1, z_2|) &=&\omega_{0,2}(z_1, z_2),\nonumber\\
\Omega^{(0)}_{0,0|2}(|w_1, w_2)&=&(x(w_1)-x(w_2))\omega_{0,2}(w_1, w_2),
\end{eqnarray}
where $\omega_{0,2}(z_1,z_2) = d z_1 d z_2/(z_1-z_2)^2$ as usual, and the Grassmann-valued meromorphic function $\gamma(z)$ on the Riemann sphere is defined by \eqref{curve2 for SEM}.

The stable differentials, with $2g-2+m+n+p\geq1$, are determined as follows:
\begin{eqnarray}
\Omega^{(0)}_{g,n|0}(J|)&=&\omega_{g,n}(J),\nonumber\\
\Omega^{(0)}_{g,n|2}(J|w_1,w_2)&=&(x(w_1)-x(w_2))\;\omega_{g,n+2}(w_1,w_2,J),\nonumber\\
\Omega^{(1)}_{g,n|1}(J|w_1)&=& \sum_{a \in \{-1, 1\}} \underset{w = a}{{\rm Res}}\ \gamma(w)\Omega_{g,n|2}^{(0)}(J|w,w_1),\nonumber\\
\Omega^{(2)}_{g,n|0}(J|)&=&\frac{1}{2} \sum_{a \in \{-1, 1\} } \underset{w = a}{{\rm Res}}\ \gamma(w)\Omega_{g,n|1}^{(1)}(J|w).\label{eq:TR for SEM}
\end{eqnarray}
\end{thm}

\subsection{Super-Gaussian Model} 
As an example, let us consider the super-Gaussian model, which is the simplest supereigenvalue model. We calculate the spectral curve and the associated Grassmann-valued function. Then by applying  Theorem \ref{main theorem} we compute a few multilinear differentials.

\subsubsection{Spectral Curve}
The super-Gaussian model is defined by the following potentials:
\begin{equation}
V(x)=\frac{1}{2}x^2,\;\;\;\;\Psi(x)=\xi_{\frac{3}{2}}x+\xi_{\frac{1}{2}}.
\end{equation}
Then we have
\begin{equation}
\mathcal{P}_{0,1|0}^{(0)}(x|)=-\frac{\partial\mathcal{F}_0^{(0)}}{\partial g_0},\;\;\;\;\mathcal{P}_{0,0|1}^{(1)}(|x)=-\frac{\partial\mathcal{F}_0^{(2)}}{\partial\xi_{\frac{1}{2}}}-\xi_{\frac{3}{2}}\frac{\partial\mathcal{F}_0^{(0)}}{\partial g_0}.\label{eq:Ps}
\end{equation}
We know from \eqref{eq:Fg0} and \eqref{Beckers formula} that
\begin{equation}
\frac{\partial\mathcal{F}_0^{(0)}}{\partial g_0}=2\frac{\partial F_0}{\partial g_0}=-2t.
\end{equation}
To calculate the last term in \eqref{eq:Ps}, we use the following trick:
\begin{eqnarray}
\frac{\partial\mathcal{F}_0^{(2)}}{\partial\xi_{\frac{1}{2}}}&=&-2\xi_{\frac{3}{2}}\frac{\partial^2}{\partial g_1^2}F_0\biggr|_{g_k=0}\nonumber\\
&=&-2\xi_{\frac{3}{2}}\frac{t^2}{N^2}\frac{\partial}{\partial g_1}\frac{1}{Z}\frac{\partial}{\partial g_1}\int \prod_{i=1}^Nd\lambda_i\Delta^2(\lambda)e^{-\frac{N}{2t}\sum_{i=1}^N(\lambda_i^2+2g_1\lambda_i)}\biggr|_{g_1=0,N=\infty}\nonumber\\
&=&-2\xi_{\frac{3}{2}}\frac{t^2}{N^2}\frac{\partial}{\partial g_1}\frac{1}{Z}\frac{\partial}{\partial g_1}e^{\frac{N^2g_1^2}{2t}}\int \prod_{i=1}^Nd\lambda'_i\Delta^2(\lambda')e^{-\frac{N}{t}\sum_{i=1}^N\lambda'^2_i}\biggr|_{g_1=0,N=\infty}\nonumber\\
&=&-2\xi_{\frac{3}{2}}\frac{t}{N}\frac{\partial}{\partial g_1}Ng_1\biggr|_{g_1=0,N=\infty}\nonumber\\
&=&-2\xi_{\frac{3}{2}}t.
\end{eqnarray}
Note that the first equality is due to \eqref{eq:beckN} with \eqref{eq:Fg0}, and we shifted $\lambda\rightarrow\lambda'=\lambda+g_1$ at the third equality. Thus, we get
\begin{equation}
\mathcal{P}_{0,1|0}^{(0)}(x|)=2t,\;\;\;\;\mathcal{P}_{0,0|1}^{(1)}(|x)=4t\xi_{\frac{3}{2}}.
\end{equation}

We set $t=1$ for simplicity. Then, the spectral curve \eqref{curve1 for SEM} and the associated Grassmann-valued polynomial \eqref{curve2 for SEM} are:
\begin{equation}
y^2=\frac{1}{4}x^2-1,\;\;\;\;y\gamma=\frac{1}{4}\left(\xi_{\frac{3}{2}}(x^2-4)+\xi_{\frac{1}{2}}x\right).
\end{equation}
The parametrization for this curve is:
\begin{equation}
x=z+\frac{1}{z},\;\;\;\;y=\frac{1}{2} \left( z-\frac{1}{z} \right),
\end{equation}
where the ramification points are at $z=\pm1$. Then $\gamma(z)$ becomes
\begin{equation}
\gamma(z)=\frac{1}{2} \left(\xi_{\frac{3}{2}}\left(z-\frac{1}{z}\right)+\xi_{\frac{1}{2}}\left(\frac{z^2+1}{z^2-1}\right) \right).
\end{equation}

\subsubsection{Topological Recursion} 
Let us explicitly calculate $\Omega^{(p)}_{g,n|m}(J|K)$ for $2g-2+n+m+p=1$ by applying Theorem \ref{main theorem}. For $\Omega^{(0)}_{0,3|0}(z,z_1,z_2|)$ and $\Omega^{(0)}_{1,1|0}(z|)$, the Eynard-Orantin topological recursion gives
\begin{eqnarray}
\Omega^{(0)}_{0,3|0}(z,z_1,z_2|)&=&\left(\frac{1}{2(z+1)^2(z_1+1)^2(z_2+1)^2}-\frac{1}{2(z-1)^2(z_1-1)^2(z_2-1)^2}\right)dzdz_1dz_2,\nonumber\\
\Omega^{(0)}_{1,1|0}(z|)&=&-\frac{z^3}{(z^2-1)^4}dz.
\end{eqnarray}
Then, from \eqref{eq:TR for SEM} we get:
\begin{align}
\Omega^{(0)}_{0,1|2}(z|w_1,w_2)=&(x(w_1)-x(w_2))\Omega^{(0)}_{0,3|0}(z,w_1,w_2|)\nonumber\\
=&\left(\frac{1}{(z+1)^2(w_1+1)^2(w_2+1)^2}-\frac{1}{(z-1)^2(w_1-1)^2(w_2-1)^2}\right)\nonumber\\
&\times \frac{(w_1-w_2)(w_1w_2-1)}{2 w_1w_2}dzdw_1dw_2,\\
\Omega^{(1)}_{0,1|1}(z|w_1)=& \sum_{a \in \{-1, 1\} }\underset{w = a}{\text{Res}}\ \gamma(w)\Omega_{0,1|2}^{(0)}(z|w,w_1)\nonumber\\
=&2 \left(\xi_{\frac{3}{2}}\frac{z}{w_1(z^2-1)^2}-\xi_{\frac{1}{2}}\frac{(z+w_1)(zw_1+1)}{(z^2-1)^2(w_1^2-1)^2}\right)dzdw_1, \\
\Omega^{(2)}_{0,1|0}(z|)=&\frac{1}{2}\sum_{a \in \{-1, 1\} } \underset{w = a}{\text{Res}}\ \gamma(w)\Omega_{0,1|1}^{(1)}(z|w)\nonumber\\
=&2 \xi_{\frac{1}{2}}\xi_{\frac{3}{2}}\frac{z}{ (z^2-1)^2}dz.
\end{align}

\begin{rem}
For the super-Gaussian model one can actually solve the super-loop equations explicitly. We verified that the differentials (for $2g-2+n+m+p \in \{-1,0,1,2\}$) obtained by Theorem \ref{main theorem} are the same (in the sense of formal expansions in $1/X$) as the correlation functions obtained by the super-loop equations, as it should be.
\end{rem}

\section{Discussion}

In this paper we showed that the Eynard-Orantin topological recursion, in conjunction with simple auxiliary equations, can be used to calculate all correlation functions of supereigenvalue models. This result is summarized in Theorem \ref{main theorem}. The geometric setup is that of a standard spectral curve, \eqref{curve1 for SEM}, and an auxiliary Grassmann-valued polynomial equation \eqref{curve2 for SEM}.

The main reason why the Eynard-Orantin topological recursion is sufficient to calculate all correlation functions is the fundamental equation \eqref{Beckers formula}, which relates supereigenvalue models to Hermitian matrix models. However, this formula is not expected to hold for multi-cut supereigenvalue models. For this purpose, one probably needs a new formalism that goes beyond the Eynard-Orantin topological recursion.

The spectral curve \eqref{curve1 for SEM} and the Grassmann-valued equation \eqref{curve2 for SEM} together can be thought as defining a ``super spectral curve'' (see for instance \cite{CHMS,CHMS2,CHJMS}). Note that these two polynomial equations are obtained from the super-loop equations without using the relation \eqref{Beckers formula} with Hermitian matrix models. Thus, one might expect that this super spectral curve provides the right initial conditions to go beyond one-cut. One would probably need to reformulate the recursive structure somehow in terms of differentials living on the super spectral curve. At this moment we do not have a clear understanding of how to proceed.

There are two other research directions that we are currently investigating. Firstly, it is interesting to ask whether one can define supereigenvalue models that satisfy super-Virasoro constraints corresponding to the super-Virasoro subalgebra in the Ramond sector, and then see whether the appropriate correlation functions can be computed recursively. We have found such a supereigenvalue partition function, and derived super-loop equations (see also \cite{CHJMS}, where such supereigenvalue models are also derived from the point of view of quantum curves).  We are currently investigating a recursive solution to these super-loop equations \cite{Osuga}. 

Secondly, recently Kontsevich and Soibelman reformulated the Eynard-Orantin topological recursion in terms of so-called ``Airy structures'' \cite{KS, ABCD, Borot}. From this viewpoint, the initial data is a sequence (finite or infinite) of quadratic differential operators that generate a Lie algebra. A natural question then is whether one can define ``super-Airy structures'', in terms of quadratic (bosonic and fermionic) differential operators that generate a super-Lie algebra. This is indeed possible, as we show in \cite{Super-Airy structure}. An open question then is whether the partition function corresponding to super-Airy structures has an interesting enumerative geometric meaning. Could it be related to enumerative geometry on the moduli-space of super-Riemann surfaces?

\appendix

\section{Derivation of \eqref{quadratic formula}}\label{sec:Derivation of the Becker's Free Energy Formula}

Here we give a proof of \eqref{quadratic formula}, which is the statement that the free energy for supereigenvalue models is at most quadratic in the Grassmann parameters. We include a proof of this fact here for completeness; it follows along similar lines to the original proof in \cite{McArthur:1993hw}.

Setting $t_s=2t$, the partition function \eqref{Z_S} of supereigenvalue models can be written as
\begin{equation}
\mathcal{Z}\overset{\text{formal}}{=}\int \prod_{i=1}^{2N}d\lambda_i\prod_{i=1}^{2N}d\theta_i\prod_{i<j}^{2N}\left(\lambda_i -\lambda_j -\theta_i\theta_j\right)e^{-\frac{N}{t}\sum_{l=1}^{2N}(V(\lambda_l)+\Psi(\lambda_l)\theta_l)}.\label{Z_S2}
\end{equation}
We will drop the ``formal'' superscript in this appendix for clarity.

We now would like to integrate over the $2N$ Grassmann variables $\theta_i$. Recall that Grassmann integrals obey
\begin{equation}
\int d\theta_k=0,\;\;\;\;\int \prod_{i=1}^{2N}d\theta_i\theta_{\sigma(1)}\cdots\theta_{\sigma(2N)}=\text{sgn}(\sigma).\label{Grassmann integral}
\end{equation}
where $\sigma \in S_{2N}$. The first equation ensures that terms with an odd number of $\xi_{k+\frac{1}{2}}$ vanish, hence the partition function is expanded as
\begin{equation}
\mathcal{Z}=\sum_{K=0}^N\mathcal{Z}^{(2K)},
\end{equation}
where the superscript denotes the order of the Grassmann couplings $\xi_{k+\frac{1}{2}}$. Note that the possible highest order of $\xi_{k+\frac{1}{2}}$ is $2N$ no matter what the degree of the Grassmann potential $\Psi(x)$ is. This is because there are only $2N$ Grassmann variables $\theta_i$ to be integrated. More precisely, we have
\begin{multline}
\mathcal{Z}^{(2K)}= \left( \frac{N}{t} \right)^{2K} \int \prod_{i=1}^{2N}d\lambda_i\prod_{i<j}^{2N}(\lambda_i-\lambda_j)e^{-\frac{N}{t}\sum_{l=1}^{2N}V(\lambda_l)}\\
\times \left(\frac{1}{(2 K)!} \int \prod_{i=1}^{2N}d\theta_i \prod_{i<j}\left(1+\frac{\theta_i\theta_j}{\lambda_j-\lambda_i}\right) \left( \sum_{l=1}^{2N} \Psi(\lambda_l) \theta_l \right)^{2 K} \right).\label{eq:Z2K}
\end{multline}
We can now evaluate the integral over the Grassmann variables $\theta_i$. It is not too difficult to see that
\begin{multline}
 \frac{1}{(2 K)!}\int \prod_{i=1}^{2N}d\theta_i \prod_{i<j}\left(1+\frac{\theta_i\theta_j}{\lambda_j-\lambda_i}\right) \left( \sum_{l=1}^{2N} \Psi(\lambda_l) \theta_l \right)^{2 K}\\
 = \frac{1}{(2K)!2^{N-K}(N-K)!}\sum_{\sigma\in S_{2N}}\text{sgn}(\sigma)\prod_{i=1}^{2K}\Psi(\lambda_{\sigma(i)})\prod_{j=K+1}^N\frac{1}{\lambda_{\sigma(2j)}-\lambda_{\sigma(2j-1)}}.\label{integral involving theta}
\end{multline}
Next, the Vandermonde determinant in \eqref{eq:Z2K} can be expressed as
\begin{equation}
\prod_{i<j}^{2N}(\lambda_i-\lambda_j)=(-1)^N\sum_{\tau\in S_{2N}}\text{sgn}(\tau)\prod_{l=1}^{2N}\lambda_l^{\tau(l)-1}.
\end{equation}
By plugging this and \eqref{integral involving theta} into \eqref{eq:Z2K}, we get:
\begin{multline}
\mathcal{Z}^{(2K)}= \left( \frac{N}{t} \right)^{2K} \frac{(-1)^N}{(2K)!2^{N-K}(N-K)!}\sum_{\tau, \sigma\in S_{2N}}\text{sgn}(\sigma)\text{sgn}(\tau)\\
\times \int \prod_{i=1}^{2N} d\lambda_i e^{-\frac{N}{t}\sum_{l=1}^{2N}V(\lambda_l)} \prod_{l=1}^{2N}\lambda_l^{\tau(l)-1}\prod_{i=1}^{2K}\Psi(\lambda_{\sigma(i)})\prod_{j=K+1}^N\frac{1}{\lambda_{\sigma(2j)}-\lambda_{\sigma(2j-1)}}.
\end{multline}
Since every $\lambda_i$ is integrated, for each permutation $\sigma \in S_n$, we can rename $\lambda_{\sigma(i)} \mapsto \lambda_i$. As a result, each term in the summation over $\sigma \in 
S_{2n}$ gives the same integral, and we get:
\begin{align}
\mathcal{Z}^{(2K)}=& \left( \frac{N}{t} \right)^{2K} \frac{(-1)^N (2N)!}{(2K)!2^{N-K}(N-K)!}\sum_{\tau \in S_{2N}}\text{sgn}(\tau)\nonumber\\
&\times \int \prod_{i=1}^{2N} d\lambda_i e^{-\frac{N}{t}\sum_{l=1}^{2N}V(\lambda_l)} \prod_{l=1}^{2N}\lambda_l^{\tau(l)-1}\prod_{i=1}^{2K}\Psi(\lambda_{i})\prod_{j=K+1}^N\frac{1}{\lambda_{2j}-\lambda_{2j-1}} \\
=&\left( \frac{N}{t} \right)^{2K} \frac{(-1)^N (2N)!}{(2K)!2^{N-K}(N-K)!}\sum_{\tau \in S_{2N}}\text{sgn}(\tau)\prod_{i=1}^{2K}\int d \lambda_i e^{-\frac{N}{t}V(\lambda_i)} \lambda_i^{\tau(i)-1}\Psi(\lambda_{i})\nonumber\\
& \times \prod_{j=K+1}^{N} \int d \lambda_{2j-1} d \lambda_{2j} e^{-\frac{N}{t}(V(\lambda_{2j-1})+V(\lambda_{2j}) )}  \frac{\lambda_{2j}^{\tau(2j)-1} \lambda_{2j-1}^{\tau(2j-1)-1}}{\lambda_{2j} - \lambda_{2j-1}}.
\end{align}
We now introduce a $2N \times 2N$ anti-symmetric matrix $A$ and a Grassmann-valued $2N$ vector $\zeta$ with components:
\begin{align}
A_{ij} =& \int d\lambda d \rho e^{- \frac{N}{t} (V(\lambda)+V(\rho))} \frac{\lambda^{i-1} \rho^{j-1}}{\lambda - \rho},\\
\zeta_i =& \frac{N}{t} \int d \lambda e^{-\frac{N}{t} V(\lambda)} \lambda^{i-1} \Psi(\lambda).
\end{align}
We can then rewrite $\mathcal{Z}^{(2K)}$ neatly as:
\begin{equation}
\mathcal{Z}^{(2K)}= \frac{(-1)^N (2N)!}{(2K)!2^{N-K}(N-K)!}\sum_{\tau \in S_{2N}}\text{sgn}(\tau) \prod_{i=1}^{2K} \zeta_{\tau(i)} \prod_{j=K+1}^N A_{\tau(2j) \tau(2j-1)}.
\end{equation}

Next, recall that the Pfaffian of a $2N \times 2N$ anti-symmetric matrix $A$ is defined by
\begin{equation}
\text{pf}(A) = \frac{(-1)^N}{2^N N!} \sum_{\sigma \in S_{2N}} \text{sgn}(\sigma) \prod_{i=1}^N A_{\sigma(2i) \sigma(2i-1)}.
\end{equation}
Thus, for $K=0$, we get directly that
\begin{equation}
\mathcal{Z}^{(0)} = (2N)!\ \text{pf}(A).
\end{equation}

To study the $K >0$ case, we need to say a little more about Gaussian Grassmann integrals. Let $M$ be an $2N \times 2N$ anti-symmetric matrix, and $\theta$ be a Grassmann-valued $2N$ vector. Then the Gaussian Grassmann integral can be evaluated as:
\begin{equation}
\int \prod_{i=1}^{2N} d \theta_i e^{- \frac{1}{2} \theta^T M \theta} =(-1)^N \text{pf}(M).
\end{equation}
This follows by expanding the exponential and integrating directly over the Grassmann variables. 

Moreover, just as for Gaussian integrals, we can also calculate shifted Gaussian Grassmann integrals. Let $M$ be an $2N \times 2N$ anti-symmetric matrix, $\theta$ be a Grassmann-valued $2N$ vector, and $\eta$ by a Grassmann-valued $2N$ vector. Then:
\begin{equation}
\int \prod_{i=1}^{2N} d \theta_i e^{- \frac{1}{2} \theta^T M \theta + \theta^T \eta} = (-1)^N \text{pf}(M) e^{\frac{1}{2} \eta^T M^{-1} \eta}.
\end{equation}
As usual, this can be obtained by completing the square inside the exponential.

With this under our belt, we can finally evaluate $\mathcal{Z}$:
\begin{align}
\mathcal{Z} =& \sum_{K=0}^N \mathcal{Z}^{(2K)} \nonumber\\
=& (-1)^N (2N)! \sum_{K=0}^N \frac{1}{(2K)!2^{N-K}(N-K)!}\sum_{\tau \in S_{2N}}\text{sgn}(\tau) \prod_{i=1}^{2K} \zeta_{\tau(i)} \prod_{j=K+1}^N A_{\tau(2j) \tau(2j-1)} \nonumber\\
=& (-1)^N (2N)! \int \prod_{i=1}^{2N} d\theta_i \prod_{j,k}^{2N} \left( 1- \frac{1}{2} \theta_j A_{j k} \theta_k \right) \prod_{l}^{2N} (1+ \theta_l \zeta_l ) \nonumber\\
=& (-1)^N (2N)! \int \prod_{i=1}^{2N} d\theta_i e^{- \frac{1}{2} \theta^T A \theta + \theta^T \zeta} \nonumber\\
=& (2N)! \text{pf}(A) e^{\frac{1}{2} \zeta^T A^{-1} \zeta} \nonumber\\
=& \mathcal{Z}^{(0)} e^{\frac{1}{2} \zeta^T A^{-1} \zeta}.
\end{align}
In other words, the free energy $\mathcal{F} = \log \mathcal{Z}$ for formal supereigenvalue models takes the form
\begin{equation}
\mathcal{F} = \log \mathcal{Z}^{(0)} + \frac{1}{2} \zeta^T A^{-1} \zeta,
\end{equation}
hence it is at most quadratic in the Grassmann coupling constants $\xi_{k+\frac{1}{2}}$.

\section{Derivation of the Loop and Super-Loop Equations}
\label{sec:Derivation of the Super Loop Equations}
In this appendix we present a derivation of the loop and super-loop equations from Virasoro and super-Virasoro constraints. An alternative derivation of the super-loop equations in terms of reparameterization of the matrix integral is discussed in \cite{Plefka:1996tt}. Note that we choose $T_2=1$ for simplicity in this section.

\subsection{Loop Equation for Hermitian Matrix Models}
\label{sec:Loop Equation for Hermitian Matrix Models}
The derivation of the loop equation starts with the following formal series
\begin{eqnarray}
\label{loop equation 1 app}
0&=&\sum_{n\geq0}\frac{1}{x^{n+1}}L_{n-1}Z\nonumber\\
&=&\frac{1}{Z}\sum_{n\geq0}\frac{1}{x^{n+1}}\left(\frac{\partial}{\partial g_{n+1}} +\sum_{k\geq0}kg_k\frac{\partial}{\partial g_{k+n-1}}+\frac{t^2}{N^2}\sum_{j=0}^{n-1}\frac{\partial}{\partial g_j}\frac{\partial}{\partial g_{n-j-1}}\right)Z,\label{derivation of loop equation}
\end{eqnarray}
where the equality holds due to the Virasoro constraints. Let us first consider the third term in \eqref{derivation of loop equation}. This term vanishes for $n=0$, hence we can shift indices:
\begin{eqnarray}
\frac{1}{Z}\sum_{n\geq0}\frac{1}{x^{n+1}}\frac{t^2}{N^2}\sum_{j=0}^{n-1}\frac{\partial}{\partial g_j}\frac{\partial}{\partial g_{n-j-1}} Z&=&\frac{1}{Z}\sum_{m\geq0}\frac{1}{x^{m+2}}\frac{t^2}{N^2}\sum_{j=0}^{m}\frac{\partial}{\partial g_j}\frac{\partial}{\partial g_{m-j}}Z\nonumber\\
&=&\frac{1}{Z}\frac{t^2}{N^2}\sum_{k,l\geq0}\frac{1}{x^{k+1}x^{l+1}}\frac{\partial}{\partial g_k}\frac{\partial}{\partial g_l}Z\nonumber\\
&=&\frac{1}{Z}\frac{t^2}{N^2}\frac{\partial}{\partial V(x)}\frac{\partial}{\partial V(x)}Z\nonumber\\
&=&\Bigl(W_1(x)\Bigr)^2+W_2(x,x).
\end{eqnarray}

On the other hand, the first two terms can be rewritten as
\begin{align}
\frac{1}{Z}\sum_{n\geq0}\frac{1}{x^{n+1}} &\left(\frac{\partial}{\partial g_{n+1}} +\sum_{k\geq0}kg_k\frac{\partial}{\partial g_{k+n-1}} \right) Z\nonumber\\
&=\left( \sum_{n \geq 0} \frac{1}{x^{n+1}} \frac{\partial}{\partial g_{n+1}} + \sum_{n,k\geq0}\frac{x^{k-1}}{x^{n+k}}kg_k\frac{\partial}{\partial g_{k+n-1}} \right)F\nonumber\\
&=\left( \sum_{n \geq 0} \frac{1}{x^{n+1}} \frac{\partial}{\partial g_{n+1}} +\sum_{m\geq0}\frac{1}{x^{m+1}}\sum_{l=0}^{m+1}x^{l-1}lg_l\frac{\partial}{\partial g_m} \right)F\nonumber\\
&=\left(x  \sum_{k \geq 1} \frac{1}{x^{k+1}} \frac{\partial}{\partial g_{k}} +\sum_{l\geq0}x^{l-1} l g_l \sum_{m \geq 0} \frac{1}{x^{m+1}}\frac{\partial}{\partial g_m}-\sum_{m\geq0}\sum_{l\geq m+2}x^{l-m-2}lg_l\frac{\partial}{\partial g_m}\right ) F\nonumber\\
&=-\frac{N}{t}W_1(x)V'(x)-\frac{\partial}{\partial g_0} F-\sum_{n\geq0}x^n\sum_{k\geq0} (n+k+2)g_{n+k+2}\frac{\partial}{\partial g_k}F,
\end{align}
where $V'(x)$ denotes the derivative of the potential with respect to $x$.
Let us denote the last two terms by $P_1(x)$, that is,
\begin{equation}
P_1(x)=-\frac{\partial}{\partial g_0} F-\sum_{n\geq0}x^n\sum_{k\geq0} (n+k+2)g_{n+k+2}\frac{\partial}{\partial g_k}F,
\end{equation}
which is a power series in $x$ (and becomes a polynomial in $x$ of degree $d-2$ if we set $g_k = 0$ for $k>d$).
Putting all this together, we obtain the loop equation \eqref{eq:loop1}:
\begin{equation}
-\frac{N}{t}V'(x)W_1(x)+P_1(x)+\Bigl(W_1(x)\Bigr)^2+W_2(x,x)=0.
\end{equation}

\subsection{Fermionic Loop Equation for Supereigenvalue Models}\label{sec:Fermionic Loop Equation for Supereigenvalue Models}
The fermionic loop equation is derived from the following formal series:
\begin{align}
0=&\frac{1}{\mathcal{Z}}\sum_{n\geq0}\frac{1}{X^{n+1}}G_{n-\frac{1}{2}}\mathcal{Z}\nonumber\\
=&\frac{1}{\mathcal{Z}}\sum_{n\geq0}\frac{1}{X^{n+1}}\left(\frac{\partial}{\partial \xi_{n+\frac{3}{2}}}+\sum_{k\geq0}\left(kg_k\frac{\partial}{\partial \xi_{n+k-\frac{1}{2}}}+\xi_{k+\frac{1}{2}}\frac{\partial}{\partial g_{k+n}}\right)+\frac{t^2}{N^2}\sum_{j=0}^{n-1}\frac{\partial}{\partial \xi_{j+\frac{1}{2}}}\frac{\partial}{\partial g_{n-j-1}}\right)\mathcal{Z},
\label{starting point of the fermionic equation}
\end{align}
where equality holds due to the super-Virasoro constraints. Let us first consider the last term. This term vanishes for $n=0$, hence we can shift indices:
\begin{align}
\frac{1}{\mathcal{Z}}\sum_{n\geq0}\frac{1}{X^{n+1}}\frac{t^2}{N^2}\sum_{j=0}^{n-1}\frac{\partial}{\partial \xi_{j+\frac{1}{2}}}\frac{\partial}{\partial g_{n-j-1}} \mathcal{Z} =& \frac{1}{\mathcal{Z}}\frac{t^2}{N^2}\sum_{m\geq0}\frac{1}{X^{m+2}}\sum_{j=0}^{m}\frac{\partial}{\partial \xi_{j+\frac{1}{2}}}\frac{\partial}{\partial g_{m-j}} \mathcal{Z} \nonumber\\
=&\frac{1}{\mathcal{Z}} \frac{t^2}{N^2} \sum_{k,l\geq 0} \frac{1}{X^{k+1}X^{l+1}}  \frac{\partial}{\partial \xi_{k+\frac{1}{2}}}\frac{\partial}{\partial g_{l}} \mathcal{Z} \nonumber\\
=&\frac{1}{\mathcal{Z}} \frac{t^2}{N^2} \frac{\partial}{\partial \Psi(X)} \frac{\partial}{\partial V(X)} \mathcal{Z}\nonumber\\
=& \mathcal{W}_{1|1}(X|X) + \mathcal{W}_{1|0}(X|) \mathcal{W}_{0|1}(|X).
\end{align}
As for the first three terms, they can be manipulated as follows:
\begin{align}
\frac{1}{\mathcal{Z}}&\sum_{n\geq0}\frac{1}{X^{n+1}}\left(\frac{\partial}{\partial \xi_{n+\frac{3}{2}}}+\sum_{k\geq0}\left(kg_k\frac{\partial}{\partial \xi_{n+k-\frac{1}{2}}}+\xi_{k+\frac{1}{2}}\frac{\partial}{\partial g_{k+n}}\right) \right) \mathcal{Z} \nonumber\\
&=X \sum_{n \geq 1} \frac{1}{X^{n+1}} \frac{\partial \mathcal{F}}{\partial \xi_{n+\frac{1}{2}}} + \sum_{k, n \geq 0} \frac{1}{X^{n+k+1}} \left(X^k (k+1) g_{k+1} \frac{\partial \mathcal{F}}{\partial \xi_{n+k+\frac{1}{2}}}+X^k \xi_{k+\frac{1}{2}}\frac{\partial \mathcal{F}}{\partial g_{k+n}} \right) \nonumber\\
&=X \sum_{n \geq 1} \frac{1}{X^{n+1}} \frac{\partial \mathcal{F}}{\partial \xi_{n+\frac{1}{2}}} + \sum_{m \geq 0} \sum_{l=0}^m \frac{1}{X^{m+1}} \left(X^l (l+1) g_{l+1} \frac{\partial \mathcal{F}}{\partial \xi_{m+\frac{1}{2}}}+X^l \xi_{l+\frac{1}{2}}\frac{\partial \mathcal{F}}{\partial g_{m}} \right) \nonumber\\
&=X \sum_{n \geq 1} \frac{1}{X^{n+1}} \frac{\partial \mathcal{F}}{\partial \xi_{n+\frac{1}{2}}} + \sum_{l\geq 0} X^l (l+1) g_{l+1}  \sum_{m \geq 0} \frac{1}{X^{m+1}} \frac{\partial \mathcal{F}}{\partial \xi_{m+\frac{1}{2}}}+ \sum_{l\geq 0}  X^l \xi_{l+\frac{1}{2}} \sum_{m \geq 0} \frac{1}{X^{m+1}} \frac{\partial \mathcal{F}}{\partial g_{m}}  \nonumber\\
& \qquad -  \sum_{m \geq 0}\sum_{l\geq m+1}  X^{l-m-1} \left((l+1) g_{l+1}  \frac{\partial \mathcal{F}}{\partial \xi_{m+\frac{1}{2}}} + \xi_{l+\frac{1}{2}} \frac{\partial \mathcal{F}}{\partial g_{m}}  \right) \nonumber\\
&= - \frac{N}{t} V'(X) \mathcal{W}_{0|1}(|X) - \frac{N}{t} \Psi(X) \mathcal{W}_{1|0}(X|) + \mathcal{P}_{0|1}(|X),\end{align}
where we defined
\begin{equation}
\mathcal{P}_{0|1}(|X) =  - \frac{\partial \mathcal{F}}{\partial \xi_{\frac{1}{2}}}-  \sum_{n \geq 0} X^{n}  \sum_{k \geq 0} \left( (n+k+2) g_{n+k+2}  \frac{\partial \mathcal{F}}{\partial \xi_{k+\frac{1}{2}}} + \xi_{n+k+\frac{3}{2}} \frac{\partial \mathcal{F}}{\partial g_{k}}  \right).
\end{equation}
Putting everything together, we find the fermionic loop equation \eqref{fermionic loop equation}:
\begin{equation}
 - \frac{N}{t} V'(X) \mathcal{W}_{0|1}(|X) - \frac{N}{t} \Psi(X) \mathcal{W}_{1|0}(X|) +\mathcal{W}_{1|1}(X|X) + \mathcal{W}_{1|0}(X|) \mathcal{W}_{0|1}(|X) + \mathcal{P}_{0|1}(|X)= 0.
\end{equation}

\subsection{Bosonic Loop Equation for Supereigenvalue Models}
\label{sec:Bosonic Loop Equation for Supereigenvalue Models}
The bosonic loop equation is derived starting from the following series:
\begin{align}
0=&\frac{1}{\mathcal{Z}}\sum_{n\geq0}\frac{1}{x^{n+1}}L_{n-1}\mathcal{Z}\nonumber\\
=&\frac{1}{\mathcal{Z}}\sum_{n\geq0}\frac{1}{x^{n+1}} \left( \frac{\partial}{\partial g_{n+1}} + \sum_{k\geq0}kg_k\frac{\partial}{\partial g_{k+n-1}}+\frac{1}{2}\left(\frac{t}{N}\right)^2\sum_{j=0}^{n-1}\frac{\partial}{\partial g_j}\frac{\partial}{\partial g_{n-j-1}} \right. \nonumber\\
&\qquad \left.+\sum_{k\geq0}\left(k+\frac{n}{2}\right)\xi_{k+\frac{1}{2}}\frac{\partial}{\partial \xi_{n+k-\frac{1}{2}}}+\frac{1}{2}\left(\frac{t}{N}\right)^2\sum_{j=0}^{n-2}\left(\frac{n-2}{2}-j\right)\frac{\partial}{\partial \xi_{j+\frac{1}{2}}}\frac{\partial}{\partial \xi_{n-j-\frac{3}{2}}}\right) \mathcal{Z}.
\label{derivation of bosonic loop equation}
\end{align}
Again, equality holds due to the super-Virasoro constraints. The first line is the same as \eqref{derivation of loop equation} except the $1/2$ in the third term. Thus it can be written as
\begin{equation}
-\frac{N}{t}V'(x)\mathcal{W}_{1|0}(x|)+\frac{1}{2} \Bigl(\mathcal{W}_{1|0}(x|)\Bigr)^2+ \frac{1}{2} \mathcal{W}_{2|0}(x,x|)-\frac{\partial \mathcal{F}}{\partial g_0} -\sum_{n\geq0}x^n\sum_{k\geq0} (n+k+2)g_{n+k+2}\frac{\partial \mathcal{F}}{\partial g_k} .
\end{equation}
We manipulate the first term in the second line of \eqref{derivation of bosonic loop equation} to get:
\begin{align}
\frac{1}{\mathcal{Z}}&\sum_{n\geq0}\frac{1}{x^{n+1}}\sum_{k\geq0}\left(k+\frac{n}{2}\right)\xi_{k+\frac{1}{2}}\frac{\partial}{\partial \xi_{n+k-\frac{1}{2}}} \mathcal{Z} \nonumber\\
=& \sum_{n \geq 0} \sum_{l \geq 0} \frac{x^l}{x^{n+l+1}} (l+1) \xi_{l+\frac{3}{2}} \frac{\partial \mathcal{F}}{\partial \xi_{n+l+\frac{1}{2}}}  -\frac{1}{2} \frac{\partial}{\partial x}   \sum_{n \geq 0} \sum_{l \geq 0} \frac{x^l}{x^{n+l+1}}  \xi_{l+\frac{1}{2}} \frac{\partial \mathcal{F}}{\partial \xi_{n+l+\frac{1}{2}}}\nonumber\\
=& \sum_{m \geq 0}  \frac{1}{x^{m+1}} \sum_{l=0}^m x^l (l+1) \xi_{l+\frac{3}{2}} \frac{\partial \mathcal{F}}{\partial \xi_{m+\frac{1}{2}}} - \frac{1}{2} \frac{\partial}{\partial x}   \sum_{m \geq 0} \frac{1}{x^{m+1}} \sum_{l =0}^m x^l  \xi_{l+\frac{1}{2}} \frac{\partial \mathcal{F}}{\partial \xi_{m+\frac{1}{2}}}\nonumber\\
=& \sum_{l \geq 0} x^l (l+1) \xi_{l+\frac{3}{2}}  \sum_{m \geq 0}  \frac{1}{x^{m+1}} \frac{\partial \mathcal{F}}{\partial \xi_{m+\frac{1}{2}}}  - \frac{1}{2} \frac{\partial}{\partial x} \sum_{l \geq 0}  x^l  \xi_{l+\frac{1}{2}}   \sum_{m \geq 0} \frac{1}{x^{m+1}}\frac{\partial \mathcal{F}}{\partial \xi_{m+\frac{1}{2}}} \nonumber\\
& \qquad - \sum_{m \geq 0} \sum_{l \geq m+1} x^{l-m-1} (l+1) \xi_{l+\frac{3}{2}} \frac{\partial \mathcal{F}}{\partial \xi_{m+\frac{1}{2}}} + \frac{1}{2} \frac{\partial}{\partial x}   \sum_{m \geq 0}\sum_{l \geq m+1} x^{l-m-1}  \xi_{l+\frac{1}{2}} \frac{\partial \mathcal{F}}{\partial \xi_{m+\frac{1}{2}}}\nonumber\\
=& - \frac{N}{t} \Psi'(x) \mathcal{W}_{0|1}(|x) + \frac{N}{2 t} \frac{\partial}{\partial x} \left( \Psi(x) \mathcal{W}_{0|1}(|x) \right) \nonumber\\
& \qquad -\frac{1}{2} \sum_{n \geq 0} x^n \sum_{k \geq 0}  \xi_{n+k+\frac{5}{2}}  \left( n+2 k+3\right) \frac{\partial \mathcal{F}}{\partial \xi_{k+\frac{1}{2}}}.
\end{align}
Finally, for the last term in \eqref{derivation of bosonic loop equation},  terms for $n=0,1$ are zero, thus we shift the index to get:
\begin{align}
\frac{1}{2 \mathcal{Z}}\left(\frac{t}{N}\right)^2& \sum_{n\geq0}\frac{1}{x^{n+1}}\sum_{j=0}^{n-2}\left(\frac{n-2}{2}-j\right)\frac{\partial}{\partial \xi_{j+\frac{1}{2}}}\frac{\partial}{\partial \xi_{n-j-\frac{3}{2}}}\mathcal{Z} \nonumber\\
=&  \frac{1}{2 \mathcal{Z} }\left(\frac{t}{N}\right)^2 \sum_{m \geq 0} \frac{1}{x^{m+3}}\sum_{j=0}^{m} \left(\frac{m}{2}-j\right)\frac{\partial}{\partial \xi_{j+\frac{1}{2}}}\frac{\partial}{\partial \xi_{m-j+\frac{1}{2}}}\mathcal{Z} \nonumber\\
=&  - \frac{1}{2 \mathcal{Z} }\left(\frac{t}{N}\right)^2 \sum_{m \geq 0} \frac{1}{x^{m+3}}\sum_{j=0}^{m} (j+1) \frac{\partial}{\partial \xi_{j+\frac{1}{2}}}\frac{\partial}{\partial \xi_{m-j+\frac{1}{2}}}\mathcal{Z}  \nonumber\\
=& - \frac{1}{2 \mathcal{Z} }\left(\frac{t}{N}\right)^2 \sum_{n \geq 0} \sum_{k \geq 0} \frac{n+1}{x^{n+2} x^{k+1}}  \frac{\partial}{\partial \xi_{n+\frac{1}{2}}}\frac{\partial}{\partial \xi_{k+\frac{1}{2}}}\mathcal{Z}  \nonumber\\
=&   \frac{1}{2 \mathcal{Z} }\left(\frac{t}{N}\right)^2 \frac{\partial}{\partial x} \left. \left( \sum_{n \geq 0} \sum_{k \geq 0} \frac{1}{x^{n+1} y^{k+1}}  \frac{\partial}{\partial \xi_{n+\frac{1}{2}}}\frac{\partial}{\partial \xi_{k+\frac{1}{2}}}\mathcal{Z}  \right) \right|_{y=x} \nonumber\\ 
=& \frac{1}{2} \mathcal{W}_{0|1}(|x)  \frac{\partial}{\partial x} \mathcal{W}_{0|1}(|x) +\frac{1}{2} \left. \frac{\partial}{\partial x} \left( \mathcal{W}_{0|2}(|x,y) \right) \right|_{y=x} \nonumber\\
\end{align}
For the second equality we used the fact that
\begin{equation}
\sum_{j=0}^{m}  \frac{\partial}{\partial \xi_{j+\frac{1}{2}}}\frac{\partial}{\partial \xi_{m-j+\frac{1}{2}}}\mathcal{Z} =0.
\end{equation}

Putting all this together, we obtain the bosonic loop equation \eqref{bosonic loop equation}:
\begin{multline}
-\frac{N}{t}V'(x)\mathcal{W}_{1|0}(x|)+\frac{1}{2} \Bigl(\mathcal{W}_{1|0}(x|)\Bigr)^2+ \frac{1}{2} \mathcal{W}_{2|0}(x,x|)- \frac{N}{2 t} \Psi'(x) \mathcal{W}_{0|1}(|x) + \frac{N}{2 t} \Psi(x)\frac{\partial}{\partial x}  \mathcal{W}_{0|1}(|x)  \\
+\frac{1}{2} \mathcal{W}_{0|1}(|x)  \frac{\partial}{\partial x} \mathcal{W}_{0|1}(|x) +\frac{1}{2} \left. \frac{\partial}{\partial x} \left( \mathcal{W}_{0|2}(|x,y) \right) \right|_{y=x} + \mathcal{P}_{1|0}(x|)= 0,
\end{multline}
where we defined
\begin{multline}
\mathcal{P}_{1|0}(x|) = -\frac{\partial  \mathcal{F}}{\partial g_0}-\sum_{n\geq0}x^n \left( \sum_{k\geq0} (n+k+2)g_{n+k+2}\frac{\partial \mathcal{F}}{\partial g_k} + \frac{1}{2} \sum_{k \geq 0}  \xi_{n+k+\frac{5}{2}}  \left( n+2 k+3\right) \frac{\partial \mathcal{F}}{\partial \xi_{k+\frac{1}{2}}} \right).
\end{multline}

\end{document}